\documentclass[useAMS,usenatbib,titlepage]{mn2e}
\usepackage{appendix}
\usepackage{amsmath}
\usepackage{txfonts}
\usepackage{graphics}
\usepackage{graphicx}
\usepackage{times}
\usepackage{placeins}
\usepackage{lastpage}
\usepackage{bm}
\usepackage{amsbsy}
\usepackage{multicol}
\usepackage{float}
\usepackage{hyperref}
\hypersetup{citecolor=blue}
\usepackage{color}
\usepackage{etoolbox}
\pretocmd{\abstractname}{\newpage}{}{}
\newcommand{\Mpcoh}{{\rm Mpc}\ h^{-1}}

\begin{document}

\title[BOSS anisotropic clustering]{The Clustering of Galaxies in the
SDSS-III Baryon Oscillation Spectroscopic Survey (BOSS): measuring growth rate and
geometry with anisotropic clustering}

\author[Samushia et al.]{
\parbox{\textwidth}{
Lado Samushia$^{1,2}$\thanks{E-mail: lado.samushia@port.ac.uk},
Beth A. Reid$^{3,4,5}$,
Martin White$^{3,5}$,
Will J. Percival$^{1}$,
Antonio J. Cuesta$^{6,7}$,
Gong-Bo Zhao$^{1,8}$,
Ashley J. Ross$^{1}$,
Marc Manera$^{9,1}$,
\'Eric Aubourg$^{10}$,
Florian Beutler$^{3}$,
Jon Brinkmann$^{11}$,
Joel R. Brownstein$^{12}$,
Kyle S. Dawson$^{12}$,
Daniel J. Eisenstein$^{13}$,
Shirley Ho$^{14}$,
Klaus Honscheid$^{15}$,
Claudia Maraston$^{1}$,
Francesco Montesano$^{16}$,
Robert C. Nichol$^{1}$,
Natalie A. Roe$^{3}$
Nicholas P. Ross$^{3,17}$,
Ariel G. S\'anchez$^{16}$,
David J. Schlegel$^{3}$,
Donald P. Schneider$^{18,19}$,
Alina Streblyanska$^{20,21}$,
Daniel Thomas$^{1}$,
Jeremy L. Tinker$^{22}$,
David A. Wake$^{23,24}$,
Benjamin A. Weaver$^{22}$,
Idit Zehavi$^{25}$
} \\
\vspace*{3pt} \\
$^{1}$ Institute of Cosmology and Gravitation, University of Portsmouth, Dennis Sciama Building, Portsmouth, P01 3FX, U.K.  \\
$^{2}$ National Abastumani Astrophysical Observatory, Ilia State University, 2A Kazbegi Ave., GE-1060 Tbilisi, Georgia  \\
$^{\star}$ lado.samushia@port.ac.uk \\
$^{3}$ Lawrence Berkeley National Laboratory, 1 Cyclotron Road, Berkeley, CA 94720, USA \\
$^{4}$ Hubble Fellow \\
$^{5}$ Departments of Physics and Astronomy, University of California, Berkeley, CA 94720, USA \\
$^{6}$ Department of Physics, Yale University, 260 Whitney Ave, New Haven, CT 06520, USA \\
$^{7}$ Institut de Ci\`encies del Cosmos, Universitat de Barcelona, IEEC-UB, Mart\'i Franqu\`es 1, E08028 Barcelona, Spain\\
$^{8}$ National Astronomy Observatories, Chinese Academy of Science, Beijing, 100012, P.R.China \\
$^{9}$ University College London, Gower Street, London WC1E 6BT, UK \\
$^{10}$ APC, Univ Paris Diderot, CNRS/IN2P3, CEA/Irfu, Obs de Paris, Sorbonne Paris Cité, F-75205, France\\
$^{11}$ Apache Point Observatory, 2001 Apache Point Road, Sunspot, NM 88349, USA\\
$^{12}$ Department of Physics and Astronomy, University of Utah, 115 S 1400 E, Salt Lake City, UT 84112, USA\\
$^{13}$ Harvard-Smithsonian Center for Astrophysics 60 Garden St. Cambridge, MA 02138 \\
$^{14}$ Department of Physics, Carnegie Mellon University, 5000 Forbes Avenue, Pittsburgh, PA 15213, USA\\
$^{15}$ Department of Physics and Center for Cosmology and Astro-Particle Physics, Ohio State University, Columbus, OH 43210, USA\\
$^{16}$ Max-Planck-Institut f\"ur Extraterrestrische Physik, Giessenbachstrasse 85748 Garching, Germany\\
$^{17}$ Department of Physics, Drexel University, 3141 Chestnut Street, Philadelphia, PA 19104, USA \\
$^{18}$ Department of Astronomy and Astrophysics, The Pennsylvania State University, University Park, PA 16802\\
$^{19}$ Institute for Gravitation and the Cosmos, The Pennsylvania State University, University Park, PA 16802\\
$^{20}$ Instituto de Astrof\'isica de Canarias (IAC), E-38200 La Laguna, Tenerife, Spain\\
$^{21}$  Universidad de La Laguna (ULL), Dept. Astrof\'isica, E-38206 La Laguna, Tenerife, Spain\\
$^{22}$ Center for Cosmology and Particle Physics, New York University, New York, NY 10003 USA\\
$^{23}$ Department of Astronomy, University of Wisconsin-Madison, 475 N. Charter Street, Madison, WI 53706\\
$^{24}$ Department of Physical Sciences, The Open University,	Milton Keynes, MK7 6AA, UK\\
$^{25}$ Department of Astronomy, Case Western Reserve University, Cleveland, OH 44106 \\
}

\date{} 
\pagerange{\pageref{firstpage}--\pageref{lastpage}}

\maketitle

\label{firstpage}

\begin{abstract}
We use the observed anisotropic clustering of galaxies in the Baryon
Oscillation Spectroscopic Survey (BOSS) Data Release 11 CMASS sample to measure the
linear growth rate of structure, the Hubble expansion rate and the comoving
distance scale. Our sample covers 8498 ${\rm deg}^2$ and encloses an effective
volume of 6 ${\rm Gpc}^3$ at an effective redshift of $\bar{z} =
0.57$. We find $f\sigma_8 = 0.441 \pm 0.044$, $H = 93.1 \pm 3.0\ {\mathrm{km}\ \mathrm{s}^{-1}
\mathrm{Mpc}^{-1}}$ and $D_{\rm A} = 1380 \pm 23\ {\rm Mpc}$ when fitting the growth and
expansion rate simultaneously. When we fix the background expansion to the one
predicted by spatially-flat $\Lambda$CDM model in agreement with recent Planck
results, we find $f\sigma_8 = 0.447 \pm 0.028$ (6 per cent accuracy).  While
our measurements are generally consistent with the predictions of $\Lambda$CDM
and General Relativity, they mildly favor models in which the strength of
gravitational interactions is weaker than what is predicted by General
Relativity. Combining our measurements with recent cosmic microwave background
data results in tight constraints on basic cosmological parameters and
deviations from the standard cosmological model. 
Separately varying these parameters, we find $w = -0.983 \pm
0.075$ (8 per cent accuracy) and $\gamma = 0.69 \pm 0.11$ (16 per cent
accuracy) for the effective equation of state of dark energy and the growth
rate index, respectively.  Both constraints are in good agreement with the 
standard model values of $w=-1$ and $\gamma = 0.554$.
\end{abstract}

\section{Introduction}

Galaxies map the distribution of the underlying dark matter field and provide
invaluable information about both the nature of dark energy (DE) and properties
of gravity \citep[see e.g.][]{Weinbergreview}.  The shape of the two-point
correlation function of the observed galaxy field, or of its Fourier-transform
the power spectrum, contains features such as baryon acoustic oscillations (BAO) and
the turn-over marking the transition between radiation dominated and matter
dominated evolutionary phases \citep{EH98,bao2}. These features can be used to place tight
constraints on relative abundances of different energy-density components of
the Universe (radiation $\rho_\gamma$, dark matter $\rho_{\rm dm}$, baryonic
matter $\rho_{\rm b}$ and DE $\rho_{\rm DE}$). Presently, these ratios
are measured to much higher accuracy in the cosmic microwave background
\citep[CMB;][]{Planck}.  Therefore, for most cosmological models these features
provide most information when used as a standard ruler. 

If the Universe is statistically isotropic and homogeneous on large-scales, the
correlation function and power spectrum should likewise be rotationally
invariant. The observed two-point statistics instead exhibit a strong
anisotropy with respect to the line-of sight (LOS) direction. Two effects are responsible for
this apparent anisotropy: the redshift-space distortions \citep[RSD;][]{RSD}
and the Alcock--Paczynski effect \citep[AP;][]{AP}.

The RSD arise in maps made from galaxies if distances are determined from
measured redshifts assuming that they are only caused by the Hubble flow.
Because of gravitational growth, the galaxies tend to infall towards high-
density regions, and flow away from low-density regions, such that the
clustering is enhanced in the LOS direction compared to the perpendicular
direction. The observed redshifts thus have a component aligned with these
flows. On large-scales where gravitational growth is linear, measuring the
relative clustering in both LOS and transverse directions leads to a
measurement of the logarithmic growth rate of structure
\begin{equation}
f(a)\sigma_8(a) = \sigma_8(a=1)\frac{dG(a)}{d\ln a}
\end{equation}
\noindent
where $a$ is a scale factor, $\sigma_8(a)$ is a measure of the
amplitude of the matter power spectrum and $G(a)$ is the linear growth function
normalized such that $G(a=1) = 1$ \citep[see][for a review of
RSD]{HamiltonReview}.

The magnitude of the large-scale velocity field traced by galaxies depends on
the nature of gravitational interactions and measured values of $f\sigma_8$ can
be used to constrain models of gravity \citep[see e.g.][]{guzzonature}. Galaxy
clustering data measures the growth at low redshifts. Combining this
information with the accurate estimates of the amplitude of matter
perturbations at $z\sim1000$  provided by CMB allows
for extremely strong constraints for deviations from the predictions of general
relativity (GR) since even small changes in the growth of structure accumulate
to a large offset over cosmic time \citep[for recent GR constraints see
e.g.][]{Zhao2012,Rapetti2013,Samushiaetal2013,Sanchez2013a,Simpson2013}.

Anisotropies are also observed due to the AP effect, which stems from the fact
that we need to convert observed angular positions and redshifts of galaxies to
physical coordinates in order to measure clustering statistics. If the
fiducial cosmology used for this mapping is different from the true cosmology
this will induce anisotropies in the measured clustering pattern even in
absence of RSD. Angular distortions are sensitive to the offset in the angular
distance $D_{\rm A}(z)$ and distortions in the LOS direction depend on the
offset in $H(z)$. Measuring the AP effect provides accurate estimates of the
angular distance and Hubble parameter and can be used to constrain properties
of DE \citep[see e.g.][]{eisenstein2007}. Measurements of both angular and
radial projected scales are usually reported in terms of the volume averaged
distance 
\begin{equation}
D_{\rm V} = \left[(1+z)^2cz\frac{D_{\rm A}^2}{H}\right]^{1/3},
\end{equation}
\noindent
and the AP-parameter
\begin{equation}
F = \frac{1+z}{c}D_{\rm A}H,
\end{equation}
In the absence of RSD, the measured correlation function monopole would be
sensitive mostly to isotropic scale dilation through $D_{\rm V}$ and the
quadrupole to anisotropic scale dilation through $F$.  Most of the information
on $D_{\rm V}$ usually comes from the most pronounced feature in the
correlation function -- the position of the BAO peak in the monopole. It is
therefore convenient to report results in terms of $D_{\rm V}/r_\mathrm{d}$ where
$r_\mathrm{d}$ is the sound horizon at the drag epoch which sets the BAO scale
\citep[for a review of BAO and AP see e.g.][]{BassettBAO}.

The RSD and AP are partially degenerate but have a different scale dependence
which makes their simultaneous measurement possible \citep{Ballinger96}.
Specifying cosmological models of background expansion or gravity helps to
further break this degeneracy \citep[e.g.][]{Samushia11}. Measuring correlation
function in different fiducial cosmological models and fitting the RSD signal
in each can help to reduce the degeneracy as well \citep{Marulli2012}.

The RSD signal within the correlation function is difficult to model because of the
significant contribution from nonlinear effects and higher order contributions
from galaxy bias. A number of recent studies have shown that many current RSD
models result in biased estimates of the growth rate \citep[see
e.g.][]{Bianchi2012,delaTorre2012, GilMarin2012}. In our work, we use the
`streaming model'-based approach developed in \citet{Reid2011} that has been
demonstrated to fit the monopole and quadrupole of the galaxy correlation
function with better than per cent level precision to scales above $25h^{-1}$
Mpc, for galaxies with bias of $b \simeq 2$.\footnote{For alternative
approaches to modelling the nonlinear effects in RSD see e.g.
\citet{Taruya2010,Okamura2011,Elia2011,Crocce2012} and \citet{Vlah2012}. For
updates to the streaming model see \citet{Wang2013}.}

Many distance-scale and RSD measurements have previously been made using
spectroscopic survey data. Recent highlights include the BAO measurements from
the 6dF Galaxy Redshift Survey \citep[6dFGRS;][]{6dFBAO}, Sloan Digital Sky
Survey II \citep[SDSS-II;][]{DR7BAO}, SDSS-III Baryon Oscillation Spectroscopic
Survey (BOSS) Data Release 9 sample \citep[DR9;][]{DR9BAO}, WiggleZ survey
\citep{WiggleZBAO} and SDSS-III BOSS DR10 and DR11 samples \citep{DR10BAO}. The
RSD signal has been measured in the 6dFGRS \citep{6dFRSD}, the SDSS-II survey
\citep{DR7RSD}, the SDSS-III BOSS DR9 data \citep{Reidetal2012} and VIMOS
Public Extragalactic Redshift Survey \citep[VIPERS;][]{VipersRSD}. Simultaneous
fits to RSD and AP parameters have been performed for the WiggleZ survey
\citep{WiggleZRSDAP}, SDSS-II data \citep{DR9RSDAP} and SDSS-III BOSS DR9 data
\citep{Reidetal2012}.

The analysis presented in this paper builds upon that of \citet{Reidetal2012},
who measured the RSD and AP simultaneously in the BOSS CMASS DR9 sample,
achieving a 15 per cent measurement of growth, 2.8 per cent measurement of angular diameter
distance, and 4.6 per cent measurement of the expansion rate at $z=0.57$.  Using these
estimates \citet{Samushiaetal2013} derived strong constraints on modified
theories of gravity (MG) and DE model parameters.  In this paper we perform a
similar analysis on the CMASS DR11 sample, which covers roughly three
times the volume of DR9.

This paper is organized as follows.  In section~\ref{sec:data} we describe the
data used in the analysis.  Section~\ref{sec:measurements} explains how the
two-dimensional correlation function is estimated from the data.
Section~\ref{sec:covariances} shows how we derive the estimates of the
covariance matrix for our measurements.  In section~\ref{sec:model} we describe
the theoretical model used to fit the data.  Section~\ref{sec:constraints}
presents and discusses our main results -- the estimates of growth rate,
distance-redshift relationship and the expansion rate from the measurements.
Section~\ref{sec:cosmoconstraints} uses these estimates to constrain parameters
in the $\Lambda$ cold dark matter ($\Lambda$CDM) model assuming GR ($\Lambda$CDM-GR) and
possible deviations from this standard model.  We conclude and discuss our
results in section~\ref{sec:conclusions}.

Our measurements require the adoption of a cosmological model in order to
convert angles and redshifts into comoving distances.  As in \citet{DR10BAO} we
adopt a spatially-flat $\Lambda$CDM cosmology with $\Omega_{\rm m} = 0.274$ and
$h=0.7$  for this purpose. For ease of comparison across analyses, we follow
\citet{DR10BAO} and also report our distance constraints relative to a model
with $\Omega_m = 0.274$, $h=0.7$, and $\Omega_{\rm b} h^2 = 0.0224$, for which
the BAO scale $r_\mathrm{d} = 149.31$ Mpc.
\section{The Data}
\label{sec:data}

The SDSS-III project \citep{SDSS3} uses a dedicated 2.5-m Sloan telescope
\citep{sdsstelescope} to perform spectroscopic follow-up of targets selected
from images made using a now-retired drift-scanning mosaic CCD camera
\citep{sdsscamera} that imaged the sky in five photometric bands
\citep{sdssbands} to a limiting magnitude of $r\simeq 22.5$.  The BOSS
\citep{boss} is the part of SDSS-III that will measure spectra for 1.5 million
galaxies and 160.000 quasars over a quarter of the sky.

We use the DR11 CMASS sample of galaxies \citep{sdsspipeline,DR10BAO,
sdssspectrograph}.  This lies in the redshift range of $0.43 < z < 0.70$ and
consists of 690826 galaxies covering 8498 square degrees (effective volume of
$6\ {\rm Gpc}^3$). Most galaxies in the sample belong to the red
sequence. About 25 per cent of them, however, would be classified as `blue'
according to traditional SDSS rest-frame colour cuts \citep[see
e.g.][]{Strateva2001}.  \citet{Ross2014} showed that there is no detectable
colour dependence of distance scale and growth rate measurements in DR10
sample.

\begin{figure}
\includegraphics[width=\linewidth]{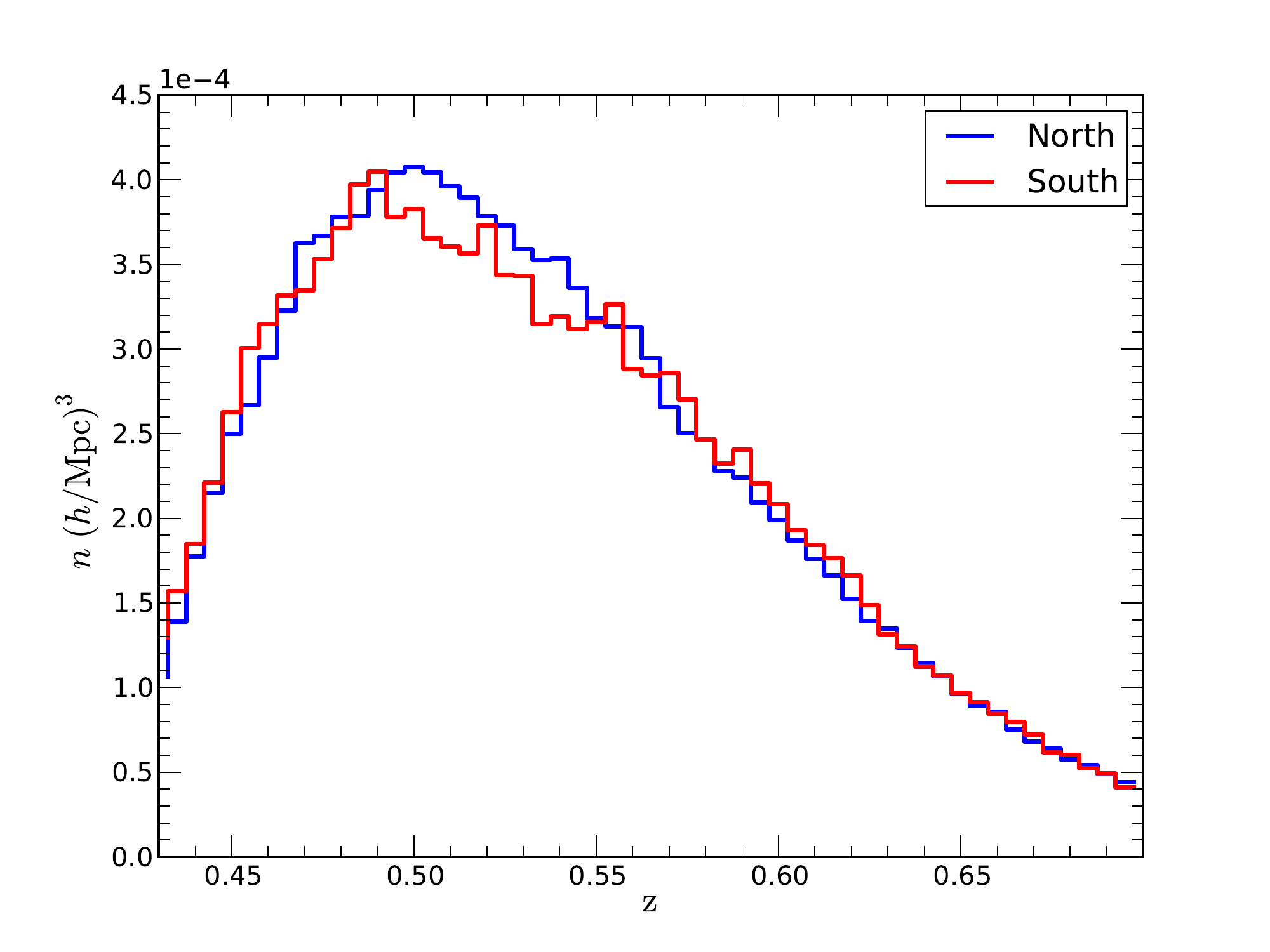}
\caption{The number density of CMASS DR11 galaxies in redshift bins of $\Delta z = 0.01$
in northern and southern Galactic hemispheres, computed assuming our fiducial cosmology.}
\label{fig:nz}
\end{figure}

Fig.~\ref{fig:nz} shows the redshift distribution of galaxies in our sample.
The number density is of the order of $10^{-4}$ peaking at $\bar{n}\simeq 4\times
10^{-4}h^3\ {\rm Mpc^{-3}}$.

\section{The measurements}
\label{sec:measurements}
We measure the correlation function of galaxies in the CMASS sample defined as
the ensemble average of the product of over-densities in the galaxy field separated by
a certain distance $\boldsymbol{r}$
\begin{equation}
\label{eq:correlation}
\xi(\boldsymbol{r}) \equiv \langle \delta_{\rm g}(\boldsymbol{r}')\delta_{\rm g}(\boldsymbol{r}' + \boldsymbol{r})\rangle.
\end{equation}
\noindent
The overdensity as a function of $\boldsymbol{r}$ is given by
\begin{equation}
\label{eq:overdensity}
\delta_{\rm g}(\boldsymbol{r}) = \frac{n_{\rm g}(\boldsymbol{r}) - \bar{n}_{\rm g}(\boldsymbol{r})}{\bar{n}_{\rm g}(\boldsymbol{r})},
\end{equation}
\noindent
where $\bar{n}_{\rm g}(\boldsymbol{r})$ is the expected average density of galaxies at a
position $\bf r$ and $n_{\rm g}(\boldsymbol{r})$ is an observed number density.

We estimate the correlation function using the Landy-Szalay minimum-variance
estimator \citep{LS}
\begin{equation}
\label{eq:LS}
\hat\xi(\Delta \boldsymbol{ r}_i) = \frac{DD(\Delta \boldsymbol{ r}_i) - 2DR(\Delta \boldsymbol{ r}_i) + RR(\Delta \boldsymbol{ r}_i)}{RR(\Delta \boldsymbol{ r}_i)},
\end{equation}
\noindent
where $DD(\Delta \boldsymbol{ r}_i)$ is the weighted number of galaxy pairs whose
separation falls within the $\Delta \boldsymbol{ r}_i$ bin, $RR(\Delta \boldsymbol{ r}_i)$ is
number of similar pairs in the random catalogue and $DR(\Delta \boldsymbol{ r}_i)$ is
the number of cross-pairs between the galaxies and the objects in the random
catalogue.

\begin{figure*}
\includegraphics[width=\linewidth]{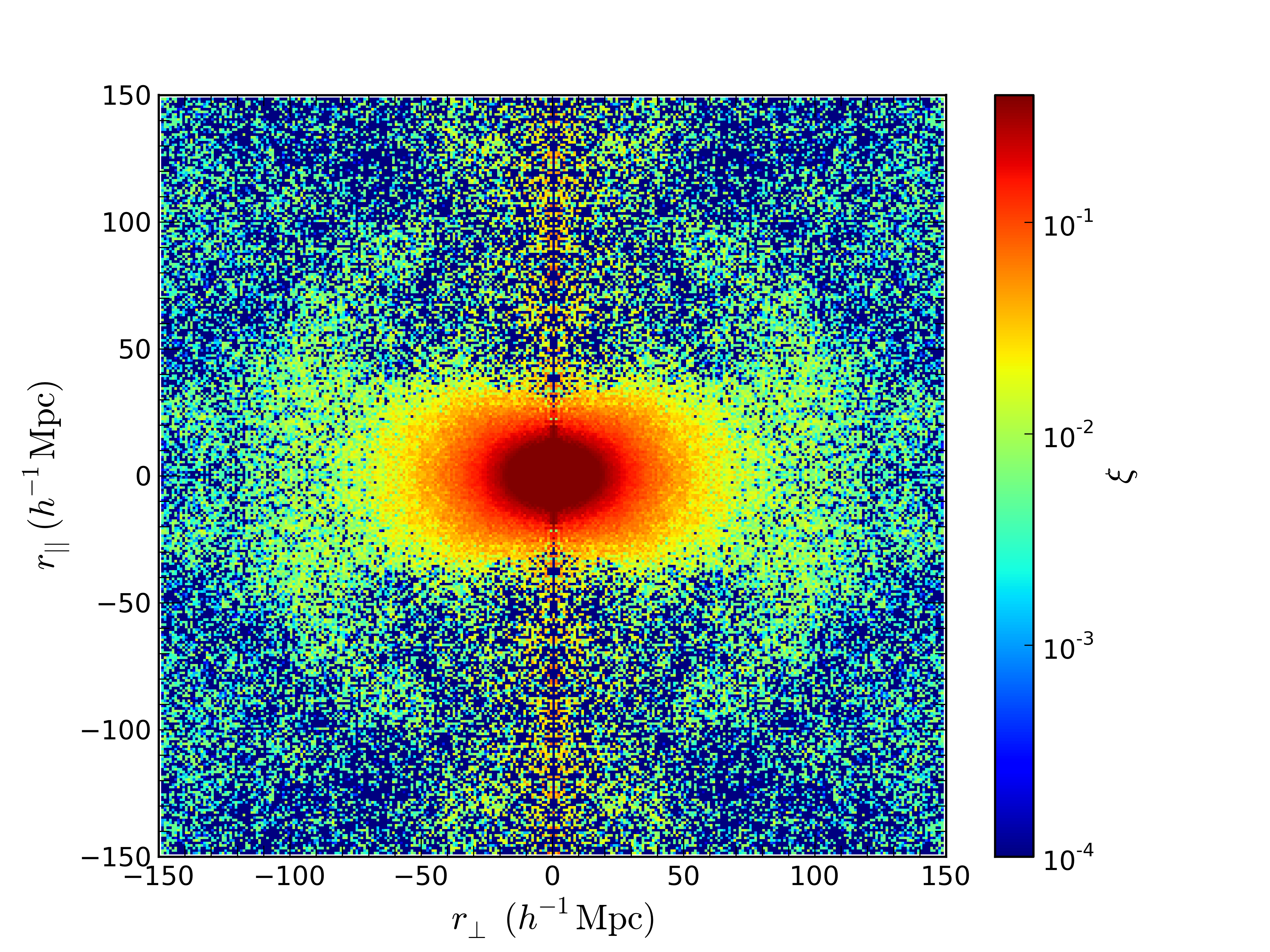} \caption{The
two-dimensional correlation function of DR11 sample measured in bins of
$1h^{-1}\times1h^{-1}$ ${\rm Mpc}^2$. We use first two Legendre multipoles of
the correlation function in our study rather than the two-dimensional
correlation function displayed here.}
\label{fig:meas2d}
\end{figure*}

Fig.~\ref{fig:meas2d} shows the two-dimensional correlation function of 
DR11 sample measured in bins of $1h^{-1}\times1h^{-1}$ ${\rm Mpc}^2$. Both the
`BAO ridge' (a ring of local maxima at approximately 100$h^{-1}$ Mpc) and the
RSD signal (LOS `squashing' of the correlation function) are detectable by
eye.

The random catalogue is constructed by populating the volume covered by
galaxies with random points with zero correlation. We use a random catalogue
that has 50 times the density of galaxies to eliminate extra uncertainty
associated with the shot noise in the random catalogue.

We weight each galaxy in the catalogue with three independent weights. First is
the Feldman--Kaiser--Peacock \citep[FKP;][]{FKP} weight $w_{\rm FKP} = 1/[1 +
\bar{n}(z)20000]$. This approach downweights galaxies in high-density regions,
achieving a balance between cosmic variance and shot-noise errors. The second
weight $w_\mathrm{sys} = w_{\rm star}w_{\rm see}$ accounts for the systematic effects
associated with both the varying stellar density
\citep[$w_\mathrm{star}$;][]{RossSystematics} and seeing variations in the
imaging catalogue used for targeting \citep[$w_\mathrm{see}$;][]{DR10BAO}. The
third weight corrects for the missed galaxies due to fibre collisions and
redshift failures using the algorithm described in \citet{DR9BAO}.  The former
is caused by the finite size of fibres that makes simultaneous measurement of
spectra of two galaxies with small angular separation impossible.  To correct
for both of these effects, we upweight each galaxy by the number of its lost
neighbours and the resulting weight is $(w_{\rm cp} + w_{\rm zf} - 1)$.  Since
these effects are statistically independent, the total weight is a product of
three $w_{\rm tot} = w_{\rm FKP}w_{\rm sys}(w_{\rm cp}+w_{\rm zf} - 1)$.  The
weight of the pair is the product of individual weights for two galaxies.
Since the stellar and close-pair effects are absent in the random catalogue we
apply only the FKP weight to them.

The observed correlation function is a function of two variables: we use
$r$, the distance between galaxies, and $\mu$, the cosine of the angle
between their connecting vector and the LOS.  The optimal choice of
binning for the correlation function measurements depends on two competing
effects.  Using small bin size retains more information, but
since we estimate covariance matrices by computing a scatter of finite number
of mock catalogues (see section~\ref{sec:covariances}), using more bins
deteriorates the precision at which the elements of the covariance matrices can be estimated.
Empirical tests performed on the mock catalogues suggest that the RSD signal is
more or less insensitive to the binning choice, while the BAO measurements are
optimal at $\sim 8h^{-1}$ Mpc \citep[for details see][]{errorspaper}. We bin
$r$ in 16 bins of $8h^{-1}$ Mpc in size in the range of $24h^{-1}\ {\rm Mpc} <
r < 152h^{-1}\ {\rm Mpc}$ and $\mu$ in 200 bins in $0 < \mu < 1$, and estimate
the correlation function on this two-dimensional grid.  The information in the
correlation function below $24\Mpcoh$ is strongly contaminated by non-linear
effects, and the scales above $152\Mpcoh$ have low signal-to-noise ratio and
contribute little information.

We compress the information in the two-dimensional correlation function by
computing the Legendre multipoles with respect to $\mu$ by approximating the
integral with a discrete sum:
\begin{eqnarray}
\hat{\xi}_\ell(r_i) &=& \frac{2\ell + 1}{2}\displaystyle\int_{-1}^{1}\mathrm{d}\mu\ \hat{\xi}(r_i, \mu) L_{\ell}(\mu) \\
  &\approx& \frac{2\ell + 1}{2}\displaystyle\sum_{k}\Delta\mu_k\ \hat{\xi}(r_i, \mu_k) L_{\ell}(\mu_k),
\label{eq:multiples}
\end{eqnarray}
\noindent
where $L_{\ell}(\mu)$ is the Legendre polynomial of the order of $\ell$.

In the subsequent analysis we only use the monopole ($\ell = 0$) and the quadrupole ($\ell = 2$)
moments. The higher order moments contain significantly less information and are more difficult to model.
\cite[For the contribution of the higher order moments see e.g.][]{Taruya2011,Kazin2012}.

The RSD signal in the measured correlation function varies within the sample
due to redshift evolution [via the redshift dependence of $f(z)\sigma_8(z)$ and
$b(z)\sigma_8(z)$]. If we keep track of the redshift of individual galaxy pairs in
equation~(\ref{eq:LS}), we effectively measure
\begin{equation}
\hat{\xi} = \frac{\displaystyle\sum \xi(z_i)w_i^2}{\displaystyle\sum w_i^2},
\end{equation} 
\noindent
where summation is over individual galaxy pairs contributing to $DD$ counts,
$\xi(z_i)$ is the correlation function at mean redshift of that galaxy pair and
$w_i^2$ is the product of the weights of individual galaxies in the ith pair.
Thus our measurement is actually a weighted redshift-averaged
correlation function.  The evolving correlation function can be expanded into
Taylor series in redshift around some value of $\bar{z}$:
\begin{equation}
\xi(z) = \xi(\bar{z}) + \left.\frac{d\xi}{dz}\right\vert_{z=\bar{z}}(\bar{z} - z) + \mathcal{O}\left[(\bar{z} - z)^2\right].
\end{equation}
\noindent
Keeping only the first-order term, we find
\begin{equation}
\label{eq:xitaylor}
\xi = \xi(\bar{z}) + \left.\frac{d\xi}{dz}\right\vert_{z=\bar{z}}\frac{\displaystyle\sum (\bar{z} - z_i)w_i^2}{\displaystyle\sum w_i^2},
\end{equation}
\noindent
and the second term disappears if we define
\begin{equation}
\label{eq:zbar}
\bar{z} = \frac{\displaystyle\sum z_i w_i^2}{\displaystyle\sum w_i^2},
\end{equation}
If the derivatives of the correlation function of second order and higher are
small, the redshift averaged correlation function is equal to the correlation
function at an `effective' redshift given by equation~(\ref{eq:zbar}).

The `effective' redshift defined in equation~(\ref{eq:zbar}) is a function of
scale. We adopt $\bar{z} = 0.57$, a value which is close to the $\bar{z}$
computed from the catalogue to better than 1 per cent precision for all
scales in the range $24h^{-1} < r < 152h^{-1}$ Mpc. We checked that in our
fiducial $\Lambda$CDM cosmology the contribution from the higher order terms
in equation~(\ref{eq:xitaylor}) are indeed small for the expected theoretical
variations in $f\sigma_8$ within the redshift range. We will therefore
interpret our estimates as measurements of the correlation function at this
effective redshift.

\begin{figure}
\includegraphics[width=\linewidth]{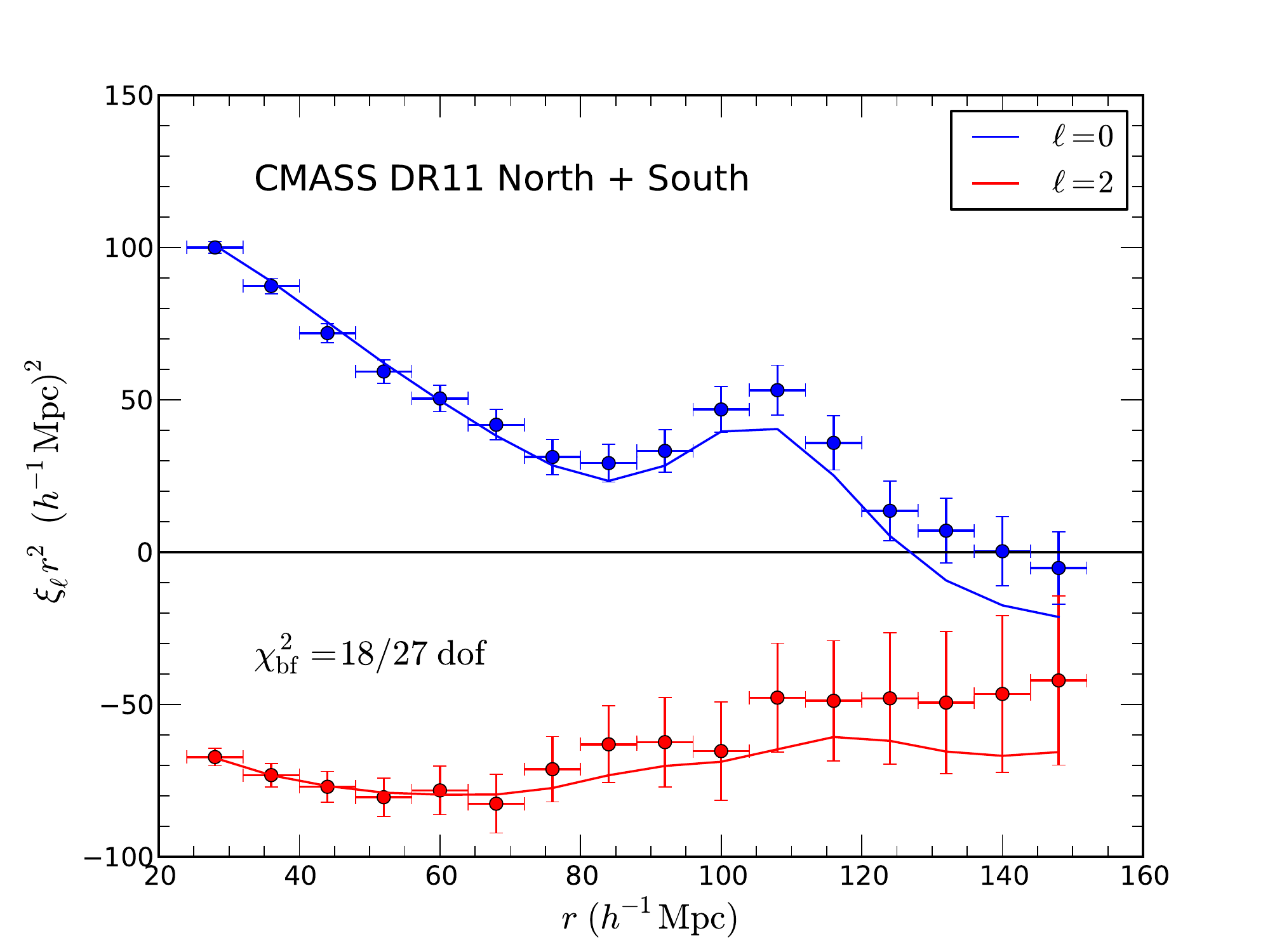}
\caption{The measured monopole and quadrupole of DR11 sample as a function of
redshift-space separation $r$. The  solid lines show predictions of our best-fitting
model with $\Omega_\mathrm{b}h^2 = 0.0222$, $\Omega_\mathrm{m}h^2 = 0.1408$,
$n_\mathrm{s} = 0.962$, $b\sigma_8=1.29$, $f\sigma_8=0.437$,
$\alpha_\bot=1.017$, $\alpha_{||}=1.001$ and $\sigma^2_{\rm FOG}=12.6$.}
\label{fig:meas}
\end{figure}

Fig.~\ref{fig:meas} shows the measured monopole and quadrupole of the CMASS
galaxies along with $1\,\sigma$ errorbars (see section~\ref{sec:covariances} for
details of the error estimation). We will use these measurements in our analysis
rather than the two-dimensional correlation function presented in
Fig.~\ref{fig:meas2d}.

\section{The covariances}
\label{sec:covariances}

To estimate the covariance matrix of our measurements we use a suite of 600 PTHalo simulations.
The simulations cover the same volume as the CMASS sample and are designed to produce a similar
bias \citep[for details of mock generation see][]{ManeraMocks}. 

\begin{figure*}
\includegraphics[width=\textwidth]{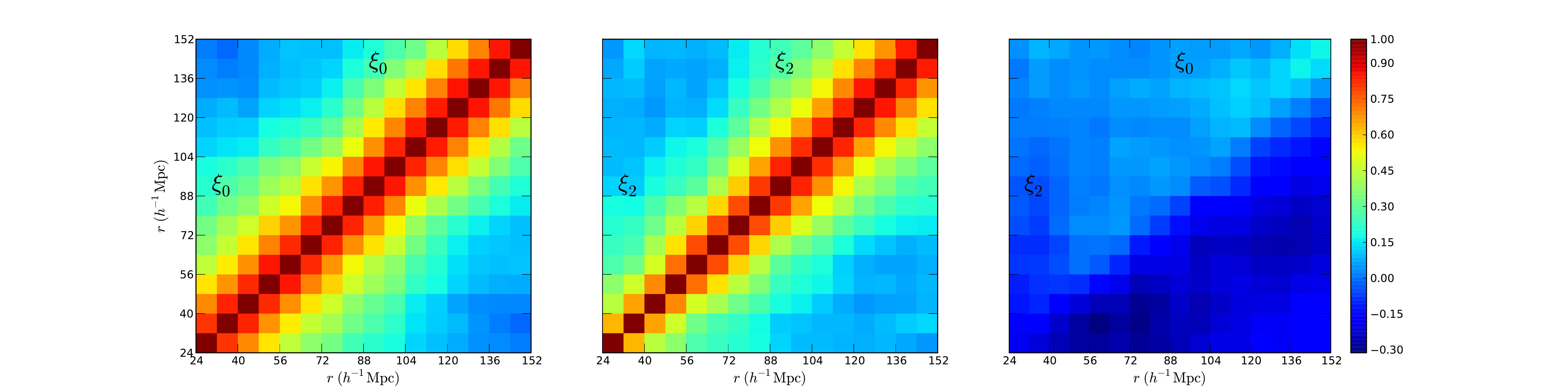}
\caption{The reduced covariance matrix ($C_{i,j}/\sqrt{C_{i,i}C_{j,j}}$) of
measured monopole and quadrupole in bins of 8$h^{-1}$ Mpc in the range
$24h^{-1} < r < 152h^{-1}$ Mpc estimated from 600 PTHalo mocks. The $\xi_\ell$
measurements in neighbouring bins are strongly correlated.}
\label{fig:redcov}
\end{figure*}

We compute the Legendre multipoles from each individual mock catalogue and estimate the covariance
matrix as 
\begin{equation}
C_{i,j}^{\ell,\ell'} = \frac{1}{N-1}\displaystyle\sum_k\left[\xi^k_\ell(r_i)-\bar{\xi}_\ell(r_i)\right]\left[\xi^k_{\ell '}(r_j)-\bar{\xi}_{\ell '}(r_j)\right],
\end{equation}
\noindent
where the sum is over individual mocks and the average multipoles
\begin{equation}
\bar{\xi}_\ell(r_i) = \frac{1}{N}\displaystyle\sum_k \xi^k_\ell(r_i)
\end{equation}
\noindent
are also computed from the mocks.
The unbiased estimator of the inverse covariance matrix is then given by
\begin{equation}
\textsf{\textbf{IC}} = \frac{N - 2 - 32}{N - 1}\textsf{\textbf{C}}^{-1},
\end{equation}
\noindent
where 32 is the number of bins used in the analysis \citep[for details
see][]{errorspaper}.  Fig.~\ref{fig:redcov} shows the reduced covariance
matrix (diagonal elements normalized to one) of our multipoles. As expected, the
measured multipoles in the neighbouring $r$-bins are strongly correlated. The
correlation between measured monopole and quadrupole at the same scale is up to
15 per cent on smaller scales.

We will compute the likelihood of theoretical models as
\begin{equation}
\label{eq:bosslik}
  \mathcal{L}\propto \exp\left(-\chi^2(\boldsymbol{ p})/2\right),
\end{equation}
\noindent
where
\begin{equation}
\label{eq:chi2}
\chi^2({\bf p}) = \displaystyle\sum_{i,j,\ell,\ell'}\left(\hat\xi_\ell(r_i)-\xi^{\rm th}_\ell(r_i, \boldsymbol{ p})\right)\widetilde{IC}_{i,j}^{\ell,\ell'}\left(\hat\xi_{\ell'}(r_j)-\xi^{\rm th}_{\ell'}(r_j, \boldsymbol{ p})\right)
\end{equation}
\noindent
$\boldsymbol{ p}$ are the set of parameters and $\xi^{\rm th}$ are the theoretical predictions for the multipoles. 
In equation~(\ref{eq:chi2}) we additionally rescale the inverse covariance matrix
\begin{eqnarray}
\label{eq:rescaleIC}
\widetilde{IC}_{i,j}^{\ell,\ell'} &=& IC_{i,j}^{\ell,\ell'} \times \frac{1 + B(n_{\rm b} - n_{\rm p})}{1 + A + B(n_{\rm p}-1)},\\
A &=& \frac{2}{(n_{\rm s}-n_{\rm b}-1)(n_{\rm s}-n_{\rm b}-4)},\\
B &=& \frac{n_{\rm s}-n_{\rm b}-2}{(n_{\rm s}-n_{\rm b}-1)(n_{\rm s}-n_{\rm b}-4)},
\end{eqnarray}
\noindent
where $n_{\rm p}$ is the length of vector $\boldsymbol{ p}$. This accounts for the
uncertainties in the determination of the inverse covariance matrix from the
finite number of catalogues \citep[for details see][]{errorspaper}.  In our
case, $n_{\rm s}=600$, $n_{\rm b}=32$ and $n_{\rm p}=5$, which results in $A =
6.25\times10^{-6}$ and $B = 1.77\times10^{-3}$.\footnote{We use $n_{\rm p}=5$
here even though the total number of fitted parameters is 8, because the three
`shape' parameters are constrained almost exclusively by the Planck
covariance matrix of equation~(\ref{eq:planckcovs}) which is not derived from our
suite of 600 PTHalo mocks and is assumed to be exact.}
The multiplicative correction factor is then $1.04$.

In approximating the likelihood by equations~(\ref{eq:bosslik}) and
(\ref{eq:chi2}), we made two assumptions: that the errors on the monopole and
quadrupole are drawn from a  multivariate Gaussian distribution
(equation~\ref{eq:bosslik}) and that the dependence of inverse covariance matrix
on free parameters is weaker than the dependence of the model
[$\mathrm{d}IC_{i,j}^{\ell,\ell'}(\boldsymbol{ p})/\mathrm{d}\boldsymbol{ p} < \mathrm{d}\xi^{\rm th}_{\ell}(\boldsymbol{
p})/\mathrm{d}\boldsymbol{ p}$ in equation~(\ref{eq:chi2})]. 

Non-linear evolution will induce non-Gaussianity. To check the validity of the
first assumption, we estimate a skewness of $\xi_\ell$ in bins of $r$ from the
600 PTHalo mocks using
\begin{equation}
\label{eq:skewness}
S_\ell(r_i) = \sqrt{600}\frac{\displaystyle\sum_k\left(\xi^k_\ell(r_i)-\bar{\xi}_\ell(r_i)\right)^3}{\left[\displaystyle\sum_k\left(\xi^k_\ell(r_i)-\bar{\xi}_\ell(r_i)\right)^2\right]^{3/2}}.
\end{equation}

\begin{figure}
\includegraphics[width=80mm]{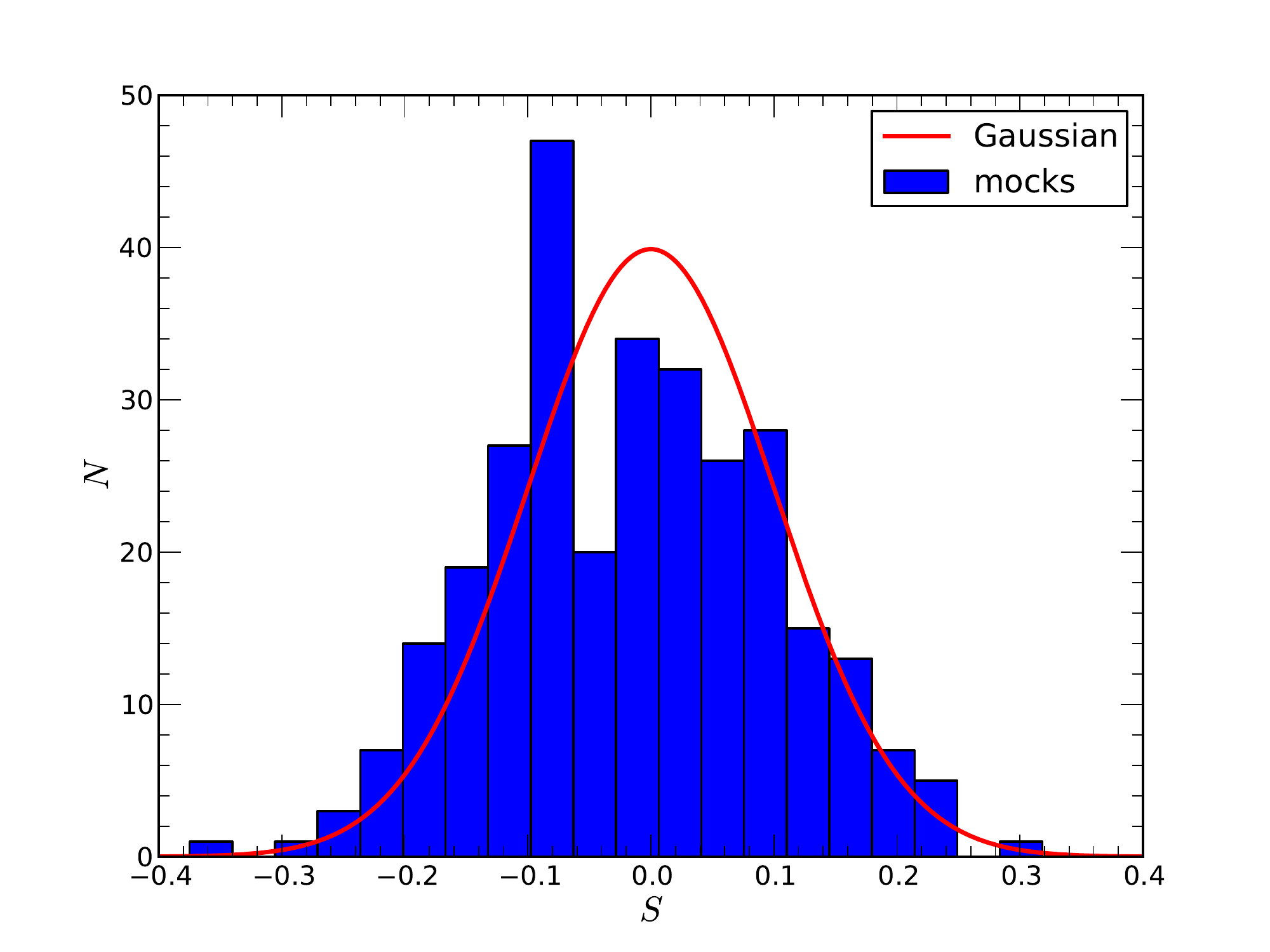}
\caption{Histogram of the skewness of monopole and quadrupole measurements
along with the expected distribution for a Gaussian variable. The empirical
variance is compatible to the expectations from a Gaussian distribution.}
\label{fig:skewness}
\end{figure}

Fig.~\ref{fig:skewness} shows a histogram of the resulting distribution of
sample skewness and the prediction made assuming that the distribution of
$\xi_\ell(r_i)$ is Gaussian. The observed distribution is consistent with the
assumption of Gaussianity; therefore, we will ignore the contribution of possible
non-Gaussian contributions to the likelihood of $\xi_\ell(r_i)$.

The validity of our second assumption is helped by the fact that the
signal-to-noise ratio is high and the mock catalogues were tuned to reproduce the
observed clustering of CMASS sample on average \citep[see][for
details]{ManeraMocks}.

\section{Theoretical model}
\label{sec:model}

\subsection{Modelling multipoles}
\label{ssec:modelmultipoles}

We use the `streaming model' to compute our theoretical template correlation
function.  Within the streaming paradigm the correlation function in redshift
space is derived by taking a real space, isotropic correlation function 
$\xi^{\rm r}(r_{||}, r_\bot)$ and
convolving it along the LOS with a probability distribution function of
the infall velocity $P$ of a galaxy pair at that separation.

\begin{equation}
\label{eq:streaming}
1 + \xi^{\rm s}(s_{||}, s_\bot) = \displaystyle\int \left[1 + \xi^{\rm r}(r_{||}, r_\bot)\right] P(s_{||} - r_{||})\mathrm{d}r_{||}
\end{equation}
\noindent
where $s_\Vert$ and $s_\bot$ are the components of a vector in the parallel and
perpendicular to the LOS direction; $r_\Vert$ and $r_\bot$ are the analogous
components in the real space.  In the plane-parallel approximation we adopt
here, $s_\bot = r_{\bot}$.  The function $P$ accounts for both quasilinear
infall motions and the random small-scale velocities \citep[`Finger-of-God' effect;][]{Jackson1972}.

Following \citet{Reid2011}, we assume
\begin{equation}
\label{eq:reidwhite}
  P(\Delta) = \frac{\exp\left(-\left[\Delta-\mu v_{\rm in}(r,\mu)\right]^2/2(\sigma_{\rm in}^2(r,\mu)+\sigma^2_{\rm FOG})\right)}{\sqrt{2\pi(\sigma_{\rm in}^2(r,\mu)+\sigma^2_{\rm FOG})}}
\end{equation}
\noindent
and compute the  $v_{\rm in}^2(r,\mu)$ and $\sigma_{\rm in}^2(r,\mu)$ values using the
standard perturbation theory, while the correlation function in the
configuration space -- $\xi^{\rm r}(\boldsymbol{ r})$ -- is computed using Lagrangian
perturbation theory \citep[see][for details]{Reid2011}. The parameter $\sigma^2_{\rm FOG}$
is an isotropic dispersion that accounts for motions of galaxies within their
local environment that are approximately uncorrelated with the large-scale
velocity field; this parameter is varied within a broad prior consistent with
the expected contribution from satellite galaxies; see \citet{Reidetal2012} for
further discussion.  

Recently, \citet{Wang2013} extended the results of convolution Lagrangian
perturbation \citep{CLPT}
and combined them with the `streaming model' to obtain accurate predictions
for the two-dimensional correlation function.
This improved model is accurate for a wider range of biases than the original
model.  For the CMASS sample, however, the original implementation of the model
in \citet{Reid2011} remains accurate enough and is used in this analysis
(see also discussion following Fig.~\ref{fig:bfmocks}).

If the real geometry of the Universe differs from the fiducial cosmology
used to compute the correlation function, this will result in the additional
distortions via the AP effect.
To account for this, we rescale the redshift-space correlation function
in equation~(\ref{eq:reidwhite}) as
\begin{equation}
\label{eq:apcorrelationfunction}
\xi^{\rm obs}(s'_\Vert, s'_\bot) = \xi^{\rm s}(\alpha_{||}s_{||}, \alpha_\bot s_\bot),
\end{equation}
\noindent
where
\begin{equation}
\label{eq:apparameters}
  \alpha_\Vert = \frac{H^{\rm fid}}{H},
  \quad \, \quad
  \alpha_\bot = \frac{D_{\rm A}}{D_{\rm A}^{\rm fid}},
\end{equation}
\noindent
and $H^{\rm fid}$ and $D_{\rm A}^{\rm fid}$ are the Hubble expansion rate and
the angular distance in the fiducial cosmology.\footnote{These parameters
were incorrectly defined in the text of \citet{Reidetal2012}, but implemented
correctly.}

The model correlation function depends on the growth rate via $v_{\rm in}$
and $\sigma_{\rm in}$.  The higher values of $f$ result in higher amplitude of
both multipoles. The dependence on the Hubble expansion rate and angular
distance arise from the AP effect and are manifested as distortions of the
multipole shapes.

Our model has been compared to N-body simulations and shown to fit the
anisotropic clustering down to scales of $\sim 25h^{-1}$ Mpc with
per cent level precision \citep{Reid2011, Reidetal2012}. To check that the PTHalo mocks
adequately describe the RSD signature in the range of scales used in the
analysis we fit our model to the mock measurements. For simplicity, we fix the
shape of the linear power spectrum to the input value and use the input
cosmology to compute radial and angular distances (this is equivalent to fixing
$\alpha_{||} = \alpha_\bot = 1$) so that the only free parameters are $f$, $b$
and $\sigma_{\rm FOG}$.  Fig.~\ref{fig:bfmocks} shows the distribution of
maximum-likelihood (ML) values of $b$ and $f$ recovered from the mock
catalogues. The systematic offset between the mean of the ML
values for $f$ and the input value of the mocks is of the order of 1 per cent, and
the scatter in the ML values
\citep[after appropriate rescaling as in][]{errorspaper} is comparable to the errors produced in
section~\ref{sec:constraints}. At least at the two-point level, this
shows that the relevant systematic effects in the PTHalos mocks are much
less than our measurement precision and can be safely ignored.

\begin{figure}
\includegraphics[width=\linewidth]{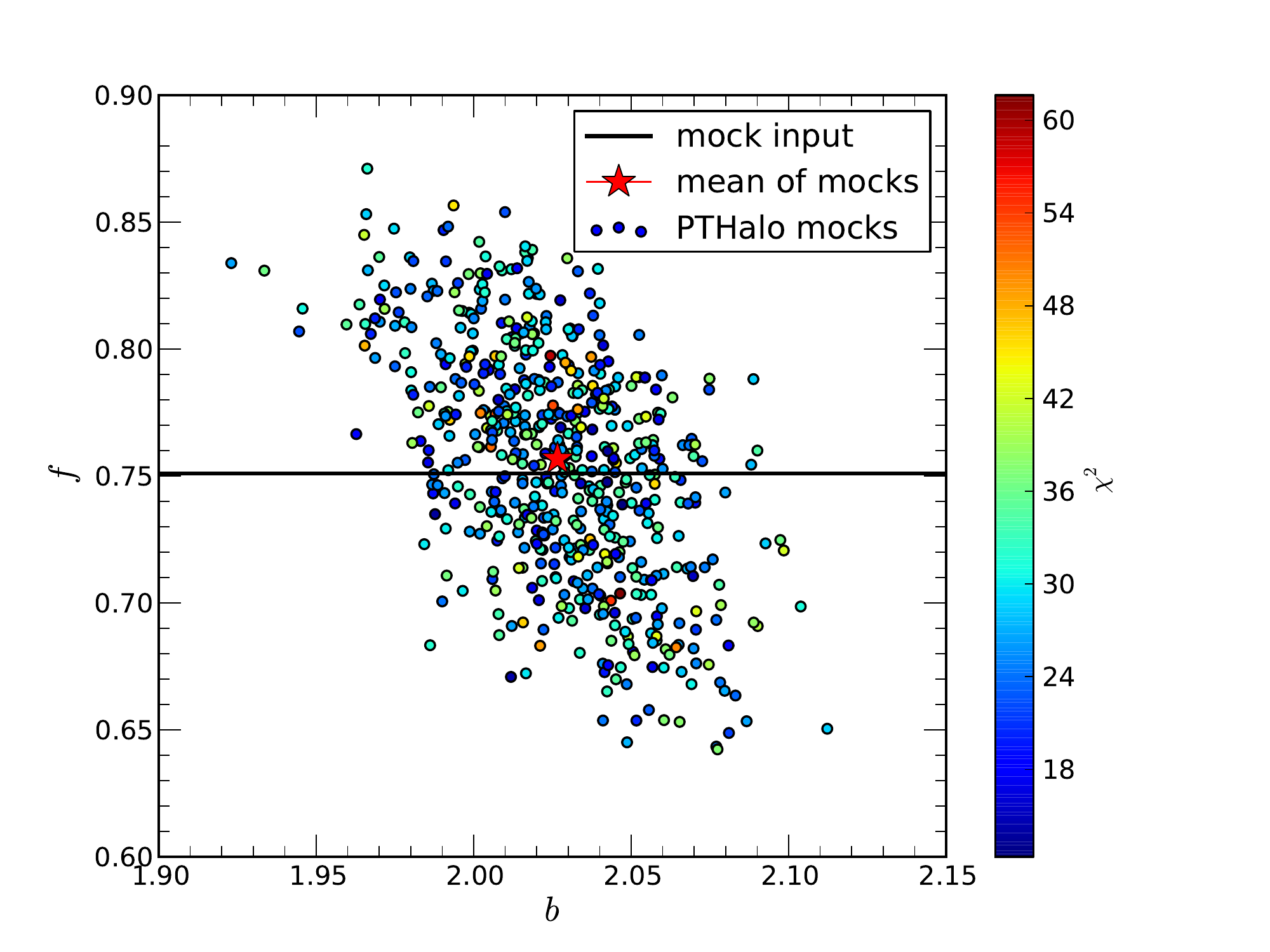}
\caption{The value of growth rate recovered from 600 PTHalo mocks.
The red star denotes the mock mean while the solid black line denotes
the input value.}
\label{fig:bfmocks}
\end{figure}

\begin{figure}
\includegraphics[width=\linewidth]{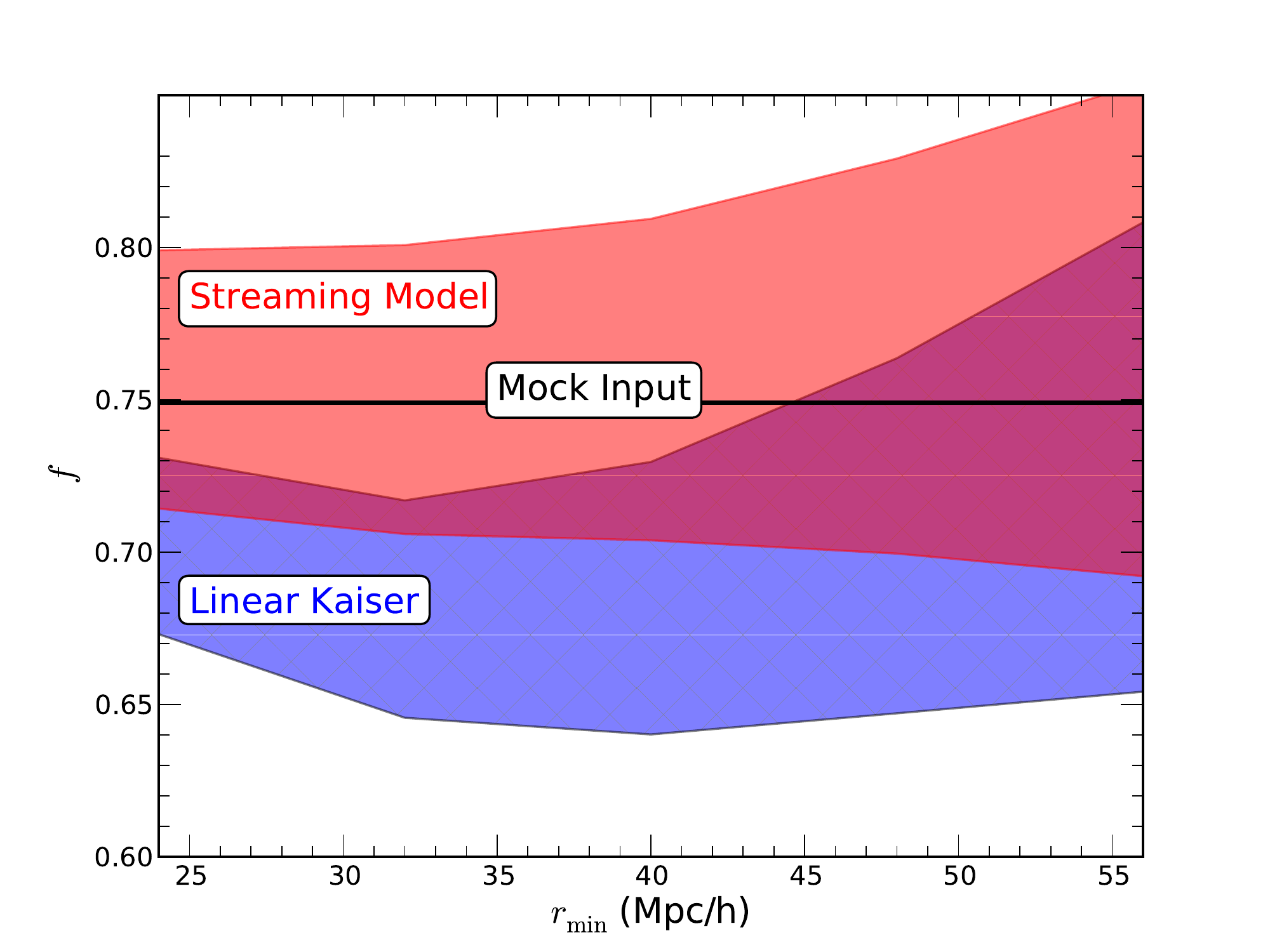}
\caption{$1\,\sigma$ interval of $f$ recovered by fitting the linear Kaiser
model (blue hatched band) and our `streaming model' (red band) to 600 PTHalo
mocks as a function of minimum scale used in the analysis.  The black solid
line denotes $f$ in the cosmology used to construct the mocks.}
\label{fig:Kaisertest}
\end{figure}

Fig.~\ref{fig:Kaisertest} shows the mean values and $1\,\sigma$ errorbars
recovered by fitting individual PTHalo mocks with the non-linear
`streaming model' as a function of minimal scale used in the analysis.
The figure also shows the results if we use a linear theory model
\citep{RSD} with no velocity dispersion nuisance parameter.
We find that for the {\it BOSS} DR11 data, one would need to fit above
$60\,h^{-1}$Mpc to get unbiased estimates of the growth rate with the linear
model.  The non-linear model resulted in unbiased fits even when scales down
to $25\,h^{-1}$Mpc were used.

The model predictions for $\xi_\ell(\boldsymbol{r})$ depend on eight parameters.  These are
parameters determining the shape of the linear correlation function
$\boldsymbol{p}_{\rm sh} = (\Omega_{\rm m}h^2, \Omega_{\rm b}h^2, n_{\rm s})$,
the bias of galaxies $b$, the linear growth rate $f$, two AP parameters
$\alpha_{||}$ and $\alpha_\bot$, and FOG velocity dispersion, $\sigma^2_{\rm
FOG}$.  In linear theory, $b$ and $f$ are completely degenerate with $\sigma_8$,
and observed clustering is only sensitive to their combination $b\sigma_8$ and
$f\sigma_8$.
Even though non-linear effects break this degeneracy, it is still present to high
degree. To compute non-linear effects on the real-space correlation function as
well as mean and variance of infall velocity, we need to specify a value of
$\sigma_8(z = 0.57)$.  In our analysis, we fix the value of
$\sigma_8(z=0.57)=0.615$, which is the best-fitting value to {\it Planck\/} data
within the $\Lambda$CDM-GR model.  We checked that model predictions do not change
significantly if we keep the values of $b\sigma_8$ and $f\sigma_8$ fixed and
vary $\sigma_8$ within $\pm3\,\sigma$ of the {\it Planck\/} constraints, and
therefore, the recovered value of $f\sigma_8$ is not sensitive to the fiducial
$\sigma_8$ used to compute non-linear effects. When combining our measurements
with \textit{Planck} data to constrain cosmological models, we do not fix a value of $\sigma_8$ and
compute it for each model accordingly (see section~\ref{sec:cosmoconstraints}).

\subsection{Modelling DE and gravity}
\label{ssec:modeldegrav}

The large-scale properties of the Universe after inflation depend on several
variables.
First, we have the relative abundances of the main energy-density constituents -- radiation
$\Omega_\gamma h^2$, baryons $\Omega_{\rm b}h^2$, dark matter $\Omega_{\rm
dm}h^2$ and DE $\Omega_{\rm de}h^2$.  We must also specify the parameters describing initial
conditions at the end of inflation -- the spectral index $n_{\rm s}$ and the
amplitude of curvature perturbations $\Delta_R^2$.  Finally, the behaviour of DE 
can be fully described by its effective equation of state (EoS) $w(z)$ if the
perturbations to DE fluid are negligible.

The energy density of radiation is determined with extremely high precision
from the temperature of microwave background and is negligible at late times.
In addition, the standard inflationary paradigm predicts the Universe to be
spatially flat to a high degree, which means that the DE energy density can be
expressed in terms of other components $\Omega_{\rm de} = 1 - \Omega_{\rm dm} -
\Omega_{\rm b}$. The parameters $\Omega_{\rm dm}$, $\Omega_{\rm b}$ and $n_{\rm s}$ are
tightly constrained by CMB data in a way that is independent of late-time
behaviour of DE and gravity. The CMB also provides a measurement of a distance, to
the last-scattering surface which depends on DE, but since it utilizes only one
integrated measurement of distance it results in highly degenerate
constraints on $w(z)$ and $\Omega_{\rm m}$ if used on its own.

The low-redshift measurements  from anisotropic galaxy clustering are strongly
complementary to CMB information. The quantities $D_{\rm A}$ and $H$ depend on
$\Omega_{\rm m}h^2 \equiv \Omega_{\rm dm}h^2 + \Omega_{\rm b}h^2$ and DE
properties, breaking degeneracies of CMB data. The $f$ in the $f\sigma_8$
measurement is sensitive to $\Omega_{\rm m}$. Relating $\sigma_8$ to $\Delta_R$
in a given model provides a strong additional test of both DE and gravity. The
fluctuations in the galaxy field ($\sigma_8$ measured by RSD) are a result of
initial fluctuations at recombination ($\Delta_R^2$ measured by CMB), and their
relationship depends on the strength of gravity and expansion of the Universe
from $z=1000$ to the redshift of the galaxy sample. 
Even small offsets from GR and $\Lambda$
are amplified and result in large offsets at low redshifts.

Independent probes of distance and expansion rate, such as measurements of
luminosity distance from supernovae Type Ia (SNIa) or direct measurements of $H$,
further enhance the cosmological constraints.

\section{Measurements}
\label{sec:constraints}

The measured monopole and quadrupole (see section~\ref{sec:measurements})
along with the covariance matrix estimated from PTHalo mocks (see
section~\ref{sec:covariances}) are fitted with the predictions of the streaming
model (see section~\ref{sec:model}) to derive constraints on the geometry and the
growth rate at an effective redshift of $\bar{z}=0.57$.

The theoretical template for the multipoles depends on the parameters $ \boldsymbol{ p} =
\left[b\sigma_8, f\sigma_8, \alpha_\Vert, \alpha_\bot, \Omega_{\rm m}h^2,
\Omega_{\rm b}h^2, n_{\rm s}, \sigma^2_\mathrm{FOG} \right]$.

Constraints on the shape of the correlation function ($\boldsymbol{ p}_{\rm sh}$) from
CMB data are significantly tighter than similar constraints obtainable from
galaxy clustering only, and these constraints are largely independent of either
the behaviour of DE at low redshifts or the nature of gravity
(i.e. the values of $\alpha_\Vert$, $\alpha_\bot$ and $f\sigma_8$ which are
of main interest here).
To exploit this fact, we multiply the likelihood in equation~(\ref{eq:bosslik}) by
a \textit{Planck} prior on this triplet
\begin{equation}
 \mathcal{L}_{\rm full} = \mathcal{L}(\boldsymbol{ p})
    \mathcal{L}_{\rm shape}(\boldsymbol{ p}_{\rm sh}),
\end{equation}
\noindent
with
\begin{align}
\chi^2(\boldsymbol{ p}_{\rm sh}) = \Delta \boldsymbol{ p}_{\rm sh}\textsf{\textbf{IC}}_\mathrm{sh} \Delta\boldsymbol{ p}_{\rm sh}^{\rm T}
\end{align}
\noindent
and the mean values of $\boldsymbol{ p}_{\rm sh}$ and the $\textsf{\textbf{IC}}_\mathrm{sh}$ are given by
\textit{Planck} temperature anisotropy data \citep{Planck}. We use the shape prior
derived from the combination of {\it Planck} temperature anisotropy data with
the {\it WMAP} low-multipole polarization likelihood which is \citep{Planck}
\begin{eqnarray}
\nonumber
\Omega_{\rm c}h^2 &=& 1.186\times 10^{-1}, \\
\label{eq:planckmeans}
\Omega_{\rm b}h^2 &=& 2.218\times 10^{-2}, \\
\nonumber
n_{\rm s} &=& 9.615\times 10^{-1},
\end{eqnarray}
\noindent
\begin{equation}
  \bordermatrix{ ~ & \Omega_{\rm c}h^2 & \Omega_{\rm b}h^2  &n_{\rm s}  \cr
  		\Omega_{\rm c}h^2 & 5.44\times 10^{5} & 6.11\times 10^{5} & 1.33\times 10^{5} \cr
		\Omega_{\rm b}h^2  & 6.11\times 10^{5} & 2.04\times 10^{7} & -2.81\times 10^{5} \cr
		n_{\rm s} & 1.33\times 10^{5} & -2.81\times 10^{5} & 6.75\times 10^{4} \cr}.
\label{eq:planckcovs}
\end{equation}

To explore this parameter space we use the nested sampling method as
implemented in the \textsc{multinest} software package
\citep{Multinest1,Multinest2}.
The free parameters of the model and their priors are listed in Table~\ref{tab:priors}.

\begin{table}
\begin{center}
\begin{tabular}{c |c | c}
\hline
Parameter & Min. value & Max. value	\\
\hline
$b\sigma_8$ & 1.0 & 1.6 \\
$f\sigma_8$  & 0.0 & 1.0 \\
$\alpha_{||}$  & 0.8  & 1.2 \\
$\alpha_\bot$ & 0.8 & 1.2 \\
$\sigma_{\rm FOG}$ & 0.0 & 50.0 \\
$\Omega_{\rm m}h^2$ & 0.08 & 0.14 \\
$\Omega_{\rm b}h^2$ & 0.018 & 0.026 \\
$n_{\rm s}$ & 0.8 & 1.2 \\
\hline
\end{tabular}
\caption{The priors on the model parameters.}
\label{tab:priors}
\end{center}
\end{table}

We have checked a posteriori that this range includes all the high likelihood
regions up to at least $5\,\sigma$ in all parameters except $\sigma_\mathrm{FOG}$
(see discussion in section~\ref{sec:comparetoothers}). The resulting
constraints on main cosmological parameters are presented in
Table~\ref{tab:constraints}.
\begin{table}
\begin{center}
\begin{tabular}{c |c | c}
\hline
Parameter & Central value & $1\,\sigma$ error	\\
\hline
$b\sigma_8$ & 1.29& 0.03 \\
$f\sigma_8$  & 0.441 & 0.043 \\
$\alpha_{||}$  & 1.006  & 0.033 \\
$\alpha_\bot$ & 1.015 & 0.017 \\
\hline
\end{tabular}
\caption{Constraints on the model parameters.}
\label{tab:constraints}
\end{center}
\end{table}

To derive constraints on DE and MG parameters we will be using the marginalised
likelihood of parameters $D_{\rm V}/r_\mathrm{d}$, $F$ and $f\sigma_8$, where 
$r_\mathrm{d}(\Omega_{\rm m}h^2, \Omega_{\rm b}h^2)$ is the sound horizon scale at
the drag epoch.  This marginalized likelihood can be approximated as a Gaussian
with mean
\begin{eqnarray}
\nonumber
D_{\rm V}/r_\mathrm{d} &=& 13.85, \\ \label{eq:mainmeans}
F &=& 0.6725, \\ \nonumber
f\sigma_8 &=& 0.4412\ .
\end{eqnarray}
\noindent
and covariance matrix
\begin{equation}
\bordermatrix{ ~ & D_{\rm V}/r_\mathrm{d} & F & f\sigma_8 \cr
  		D_{\rm V}/r_\mathrm{d} & 2.88\times 10^{-2} & -9.67\times 10^{-4} & -4.46\times 10^{-4} \cr
		F & -9.67\times 10^{-4} & 7.98\times 10^{-4} & 9.70\times 10^{-4} \cr
		f\sigma_8 & -4.46\times 10^{-4} & 9.70\times 10^{-4} & 1.89\times 10^{-3} \cr}.
\label{eq:maincovs}
\end{equation}
Equations~(\ref{eq:mainmeans}) and (\ref{eq:maincovs}) use values of
$r_\mathrm{d}=r_s(z_{\rm d})$ derived by numerically integrating the recombination
equations and integrating the sound speed up to the drag epoch.  These values
are related to the results derived from commonly used fitting formula of
\citet{EH98} adjusted by a factor of $r_\mathrm{d}^{\rm EH}/r_\mathrm{d}=1.026$.  This ratio is
independent of cosmology for a wide range of conventional cosmological models
\citep[see e.g.][]{Mehta2012}.

\begin{figure*}
\includegraphics[width=\textwidth]{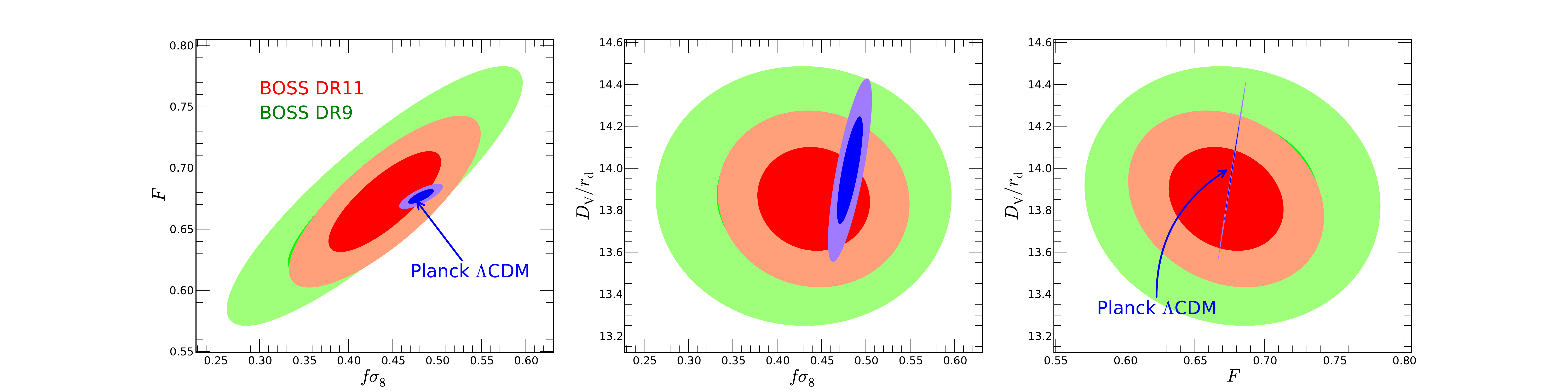}
\caption{Posterior likelihood of parameters $D_{\rm V}/r_\mathrm{d}$, $F$ and
$f\sigma_8$ from {\it BOSS} DR11 (red contours) and {\it BOSS} DR9 (green
contours) data, along with expectations from Planck data within standard
$\Lambda$CDM-GR models (blue contours). All estimates are mutually consistent.}
\label{fig:dvFfs8}
\end{figure*}

Fig.~\ref{fig:dvFfs8} shows the constraints on main cosmological parameters
compared to the expectations from the \textit{Planck} data within standard
$\Lambda$CDM-GR models along with DR9 results from \citet{Reidetal2012}.
The DR11 results are in a good agreement with the {\it Planck}
predictions; the $\chi^2$ difference between them is 1.6 for 3 degrees of
freedom. 

Equations~(\ref{eq:mainmeans}) and (\ref{eq:maincovs}) represent the main
results of our work and will be used later to constrain models of DE and MG
(see section~\ref{sec:cosmoconstraints}).

\subsection{Comparison to other similar measurements}
\label{sec:comparetoothers}

The companion papers, \citet{DR10BAO}, \citet{Beutler2013},
\citet{Sanchez2013b} and \citet{Chuang2013} use the same CMASS DR11 data to
constrain the distance--redshift relation at $z=0.57$.  

\begin{figure}
\includegraphics[width=\linewidth]{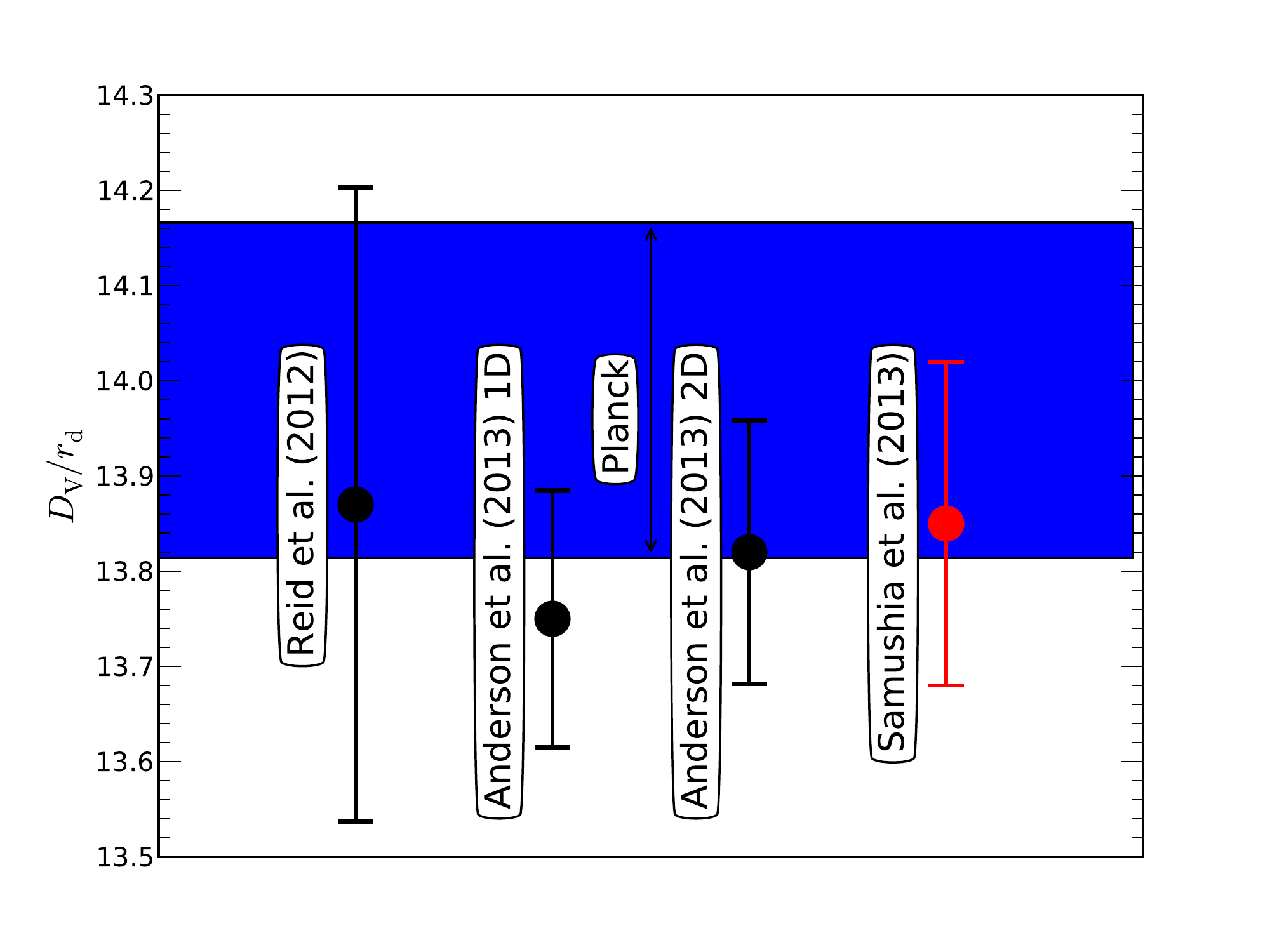}
\caption{Various estimates of $D_{\rm V}/r_\mathrm{d}$ from CMASS DR9 and 
DR11 data sets. The blue band corresponds to 1$\sigma$ uncertainty in {\it
Planck} prediction assuming $\Lambda$CDM. All measurements are mutually consistent.}
\label{fig:DVcompare}
\end{figure}

Fig.~\ref{fig:DVcompare} shows our measurement of distance along with the
result from BAO only fits and previous similar measurements and {\it Planck}
predictions for spatially-flat $\Lambda$CDM model.

In Fig.~\ref{fig:DVcompare}, the label 1D refers to the result derived by
fitting the monopole of the correlation function only, while the label 2D
refers to the result derived from the fit to the monopole and the quadrupole of
the correlation function \citep[see][for details]{DR10BAO}.
differ from our analysis in two important aspects. They apply
`reconstruction' to the measured galaxy distribution to partially remove the
nonlinear smearing of the BAO feature, and marginalize over the broad-band shape
of the correlation function, so that the estimate of the distance comes from
the BAO peak feature alone. 

\citet{Beutler2013} and \citet{Chuang2013} measured the distance--redshift
relationship using the Legandre moments of power spectrum and correlation
function, respectively. \citet{Beutler2013} perform their analysis in Fourier
space.  The \citet{Chuang2013} analysis is in configuration space but uses a
different range of scales and theoretical model than our work. Despite
differences in the applied methodology, the estimates are consistent within
1$\sigma$ error bars. 

\begin{figure}
\includegraphics[width=\linewidth]{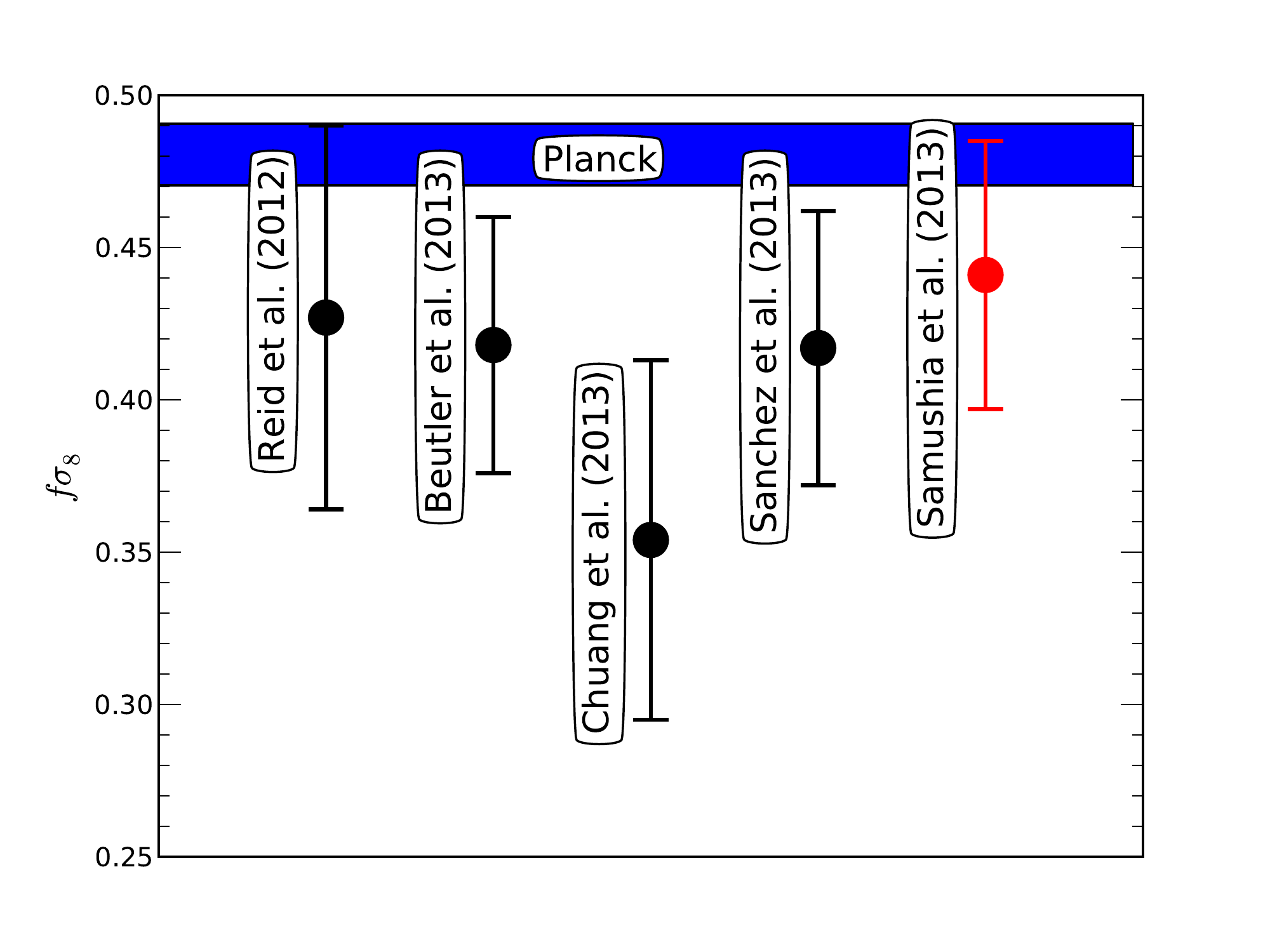} 
\caption{Various estimates
of $f\sigma_8$ from CMASS DR9 and DR11. The blue band corresponds
to $1\,\sigma$ uncertainty in {\it Planck} prediction assuming $\Lambda$CDM-GR.
Clustering measurements are mutually consistent and are lower than the CMB
prediction.}
\label{fig:comparefs8}
\end{figure}

The growth rate, $f\sigma_8$, has also been measured in the same redshift bin
by \citet[][DR11]{Beutler2013}, \citet[][DR9]{Reidetal2012}, \citet[][DR11]{Chuang2013}
and \citet{Sanchez2013b}. The comparison of results is presented in
Fig.~\ref{fig:comparefs8}. In the \citet{Sanchez2013b} analysis, $f\sigma_8$
is a derived parameter computed by combining CMASS data with {\it Planck}
assuming $\Lambda$CDM model; their estimate is perfectly consistent with ours.
The \citet{Reidetal2012} analysis is similar in the range of scales and
theoretical modelling to the current paper, but performed on DR9 data set. All
measurements are consistent with each other and are somewhat lower than the
\textit{Planck} $\Lambda$CDM-GR expectations.

\subsubsection{Comparison with our DR9 measurements}

The fitting methodology adopted in this paper is identical to that used in our
DR9 analysis \citep{Reidetal2012}, but some of the priors have been updated.
We adopt a prior on the linear matter power spectrum shape from {\it Planck}
rather than \textit{WMAP7}; {\it Planck} has substantially smaller errors, and so we
expect the marginalization over the $P(k)$ to contribute negligibly to our
error budget in DR11.
We also adopted a slightly more conservative top-hat prior on
$\sigma^2_\mathrm{FOG}$, by increasing the allowed range from 0 -- 40${\rm Mpc}^2$ to
0 -- 50${\rm Mpc}^2$, as the large-scale clustering data alone can-not well constrain
this dispersion term; we have checked that this change of prior range
does not affect our best-fitting parameter values significantly.

The effective area of DR11 is a factor of 2.5 larger than DR9; in the limit of
negligible boundary effects, we would expect the covariance matrix on
$D_\mathrm{V}/r_\mathrm{d}, F$, and $f\sigma_8$ to be reduced by the same factor.  A
direct comparison indicates agreement at the $\sim 15\%$ level on the
diagonals, with DR11 errors slightly larger than expected and with different
off-diagonal structure.  When projected on to $f\sigma_8$ (at fixed
$D_\mathrm{V}/r_\mathrm{d}$ and $F$), which is the relevant case for the
modified gravity constraints we present, our error in DR9 was 0.033 and is
0.028 in DR11, while we would have expected 0.021 from the effective volumes.
This situation arises because, as we showed in table 2 of \citet{Reidetal2012},
the prior on $\sigma^2_\mathrm{FOG}$ reduces the uncertainty on $f\sigma_8$ in the
fixed geometry case substantially.  The statistical errors have shrunk
significantly in DR11, but we did not assume better prior knowledge on
$\sigma^2_\mathrm{FOG}$.

Fisher matrix analysis suggests that if $\sigma^2_\mathrm{FOG}$ parameter were
perfectly known, the $f\sigma_8$ error would be reduced to 0.017 when the
geometric and power spectrum parameters are held fixed.\footnote{For an update
on the small-scale $\sigma^2_{FOG}$ estimate and its effect on $f\sigma_8$
measurement see \citet{Reidprep}.}

In DR11 we obtain higher values for $D_{\rm V}/r_{\rm s}$ and $f\sigma_8$,
which brings us slightly closer to the values predicted by {\it Planck}. The
$\chi^2$ offset between DR11 and DR9 results is just 0.3 per 3 degrees of
freedom. 

\subsection{Constraints from monopole and quadrupole Separately}

\begin{figure}
\includegraphics[width=\linewidth]{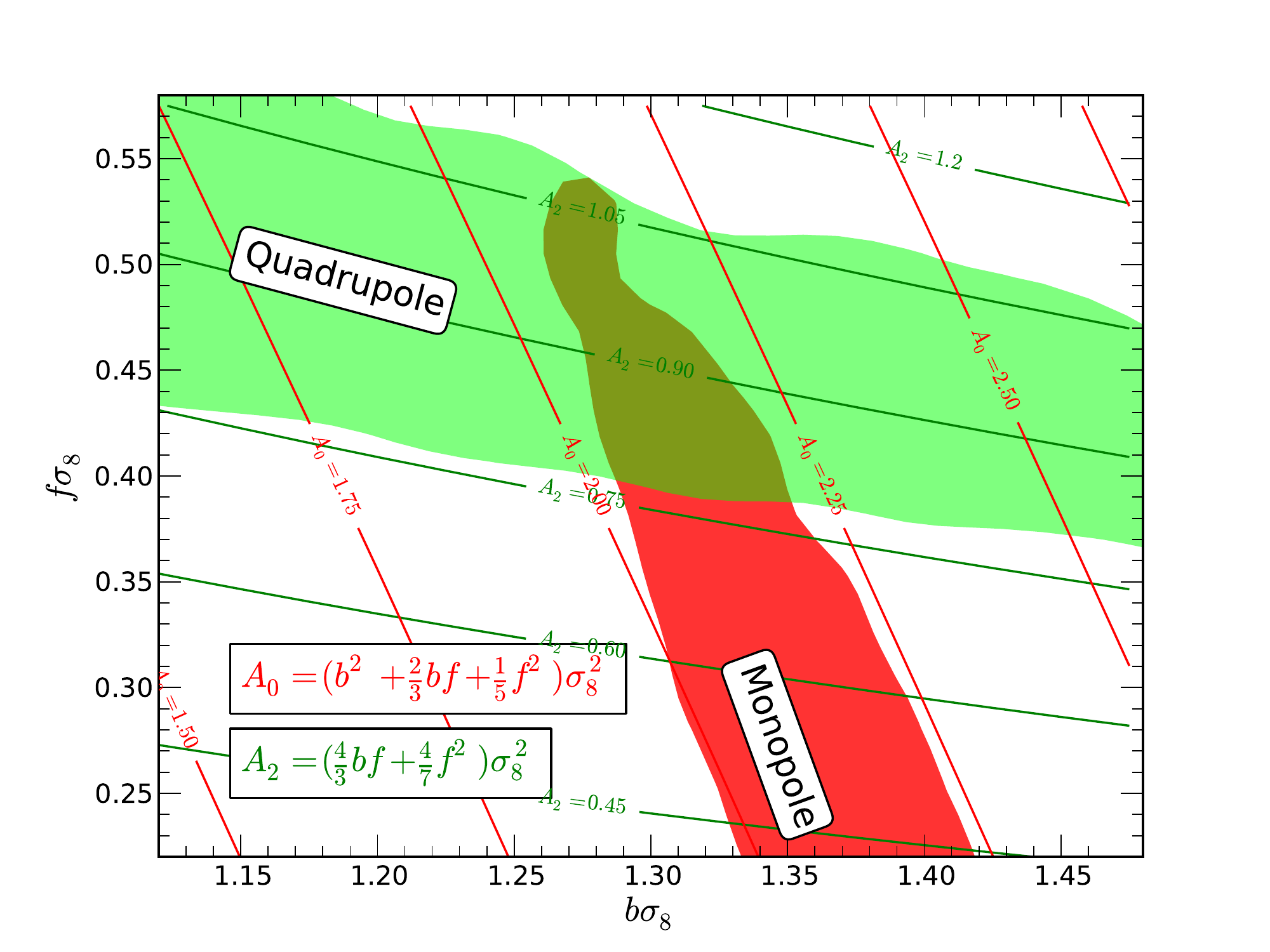}
\caption{Constraints on $b\sigma_8$ and $f\sigma_8$ from monopole and quadrupole
separately. Solid lines show expected directions of the principal components based
on predictions of the linear theory.}
\label{fig:bf}
\end{figure}

To determine the separate contribution of monopole and quadrupole we perform the same fit
to each individually. The monopole and quadrupole measurements on 
their own are unable to break the degeneracy between $b\sigma_8$ and $f\sigma_8$
and can only constrain combinations of the two. Fig.~\ref{fig:bf} shows
the constrains in $b\sigma_8$ -- $f\sigma_8$ derived from the two multipoles.
The solid lines show the expected degeneracy directions based on the linear theory
predictions.

\begin{figure}
\includegraphics[width=\linewidth]{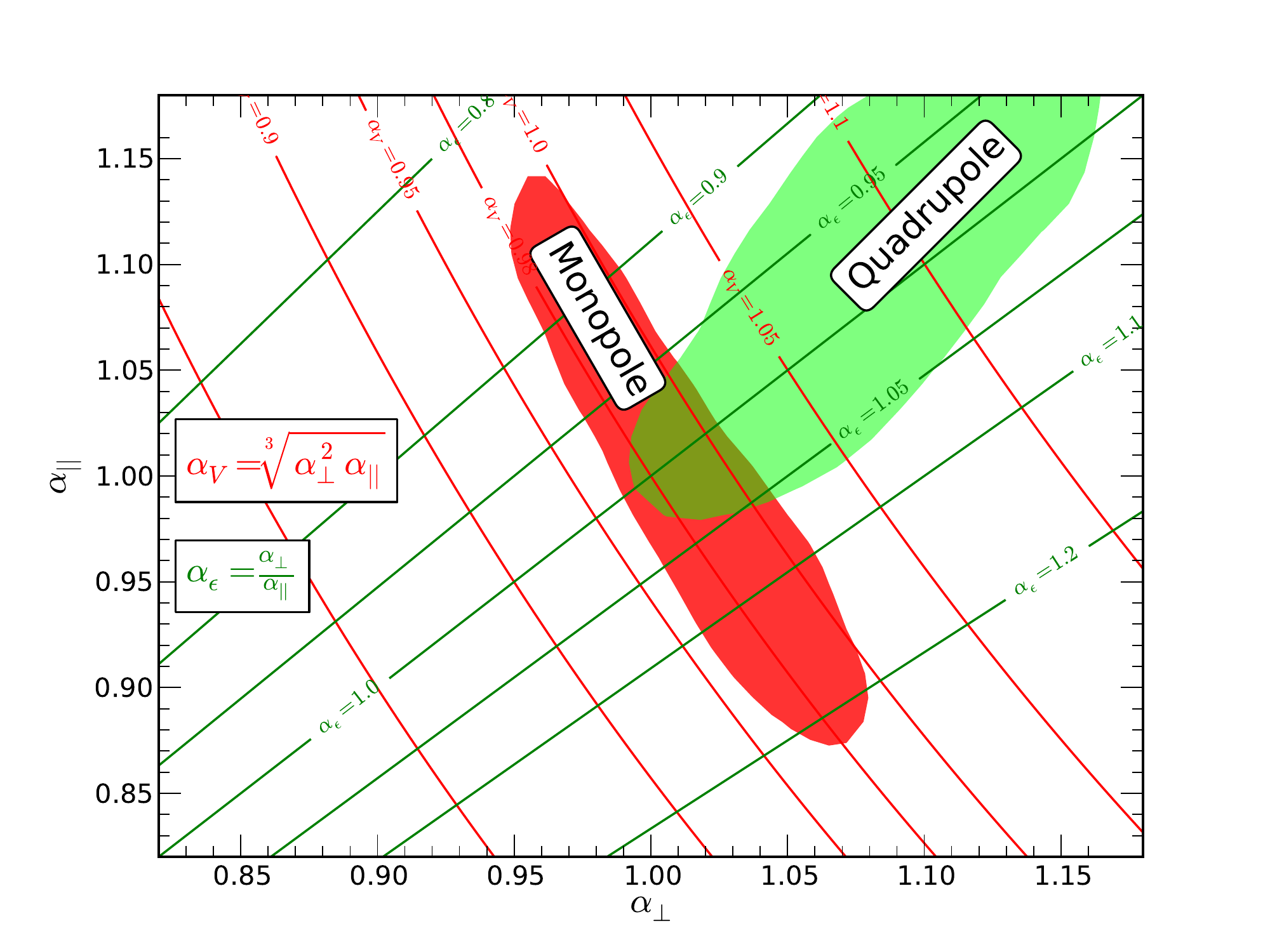}
\caption{Constraints on $\alpha_{||}$ and $\alpha_\bot$ from monopole and quadrupole
separately. Solid lines show expected directions of the principal components based
on predictions of the linear theory.}
\label{fig:alphaepsilon}
\end{figure}

The quadrupole best constrains $A_2 = (4/3bf + 4/7f^2)\sigma_8^2$, as
expected from the linear theory. The amplitude constraints from monopole are
collinear to the combination $A_0 = (b^2 + 2/3bf + 1/5f^2)\sigma_8^2$, also as expected from
linear theory.

The AP parameters $\alpha_{||}$ and $\alpha_\bot$ show a qualitatively similar
picture. Individual multipoles can only constrain certain combinations of parameters.
Fig.~\ref{fig:alphaepsilon} presents constraints in the $\alpha_{||}$ --
$\alpha_\bot$ plane from the monopole and quadrupole separately. The solid
lines show the expected degeneracy directions based on the linear theory
predictions.

The principal component of the monopole constraint coincides with $\alpha_{\rm
V} = \sqrt[3]{\alpha_\bot^2\alpha_{||}}$ as expected from linear theory. The
principal component of the quadrupole constraint is slightly tilted from
the $\alpha_\epsilon\equiv\alpha_\bot/\alpha_{||}={\rm const}$ direction; 
$\alpha_{\rm V} = 1.011 \pm 0.013$ from the monopole and $\alpha_\epsilon =
0.988 \pm 0.091$ from the quadrupole.

\subsection{Separate fits to growth and AP}

We next fit the monopole and quadrupole for the growth factor and AP parameters
separately.  First, we assume that the background expansion follows the
predictions of spatially flat $\Lambda$CDM but allows the growth rate to be a
free parameter. In this case, the parameters $\alpha_\Vert$ and $\alpha_\bot$
can be computed from $\Omega_{\rm m}$ and $H_0$.  For this model, where the background
expansion is assumed to be following the $\Lambda$CDM predictions, we find
$f\sigma_8 = 0.447 \pm 0.028$ and $b\sigma_8 = 1.26 \pm 0.02$. The constraint
on growth improves to 6 per cent (from 10 per cent) and is perfectly consistent
with the result of our more general fit.

Next, we assume that the growth rate follows the predictions of $\Lambda$CDM-GR,
but let the expansion rate and the distance--redshift relation vary. In this
case $f\sigma_8$ is computed from $\Omega_{\rm m}$ but the $\alpha_\Vert$ and
$\alpha_\bot$ are free parameters. For this fit, we obtain $\alpha_\Vert = 0.992 \pm
0.023$ and $\alpha_\bot = 1.021 \pm 0.013$. Constraints on $\alpha_\Vert$ move to
a lower value and tighten to 2 per cent (from 3 per cent), while constraints on
$\alpha_\bot$ move to a higher value and tighten to 1 per cent (from 2 per
cent).

\subsection{Contribution from small scales}

\begin{figure}
\includegraphics[width=\linewidth]{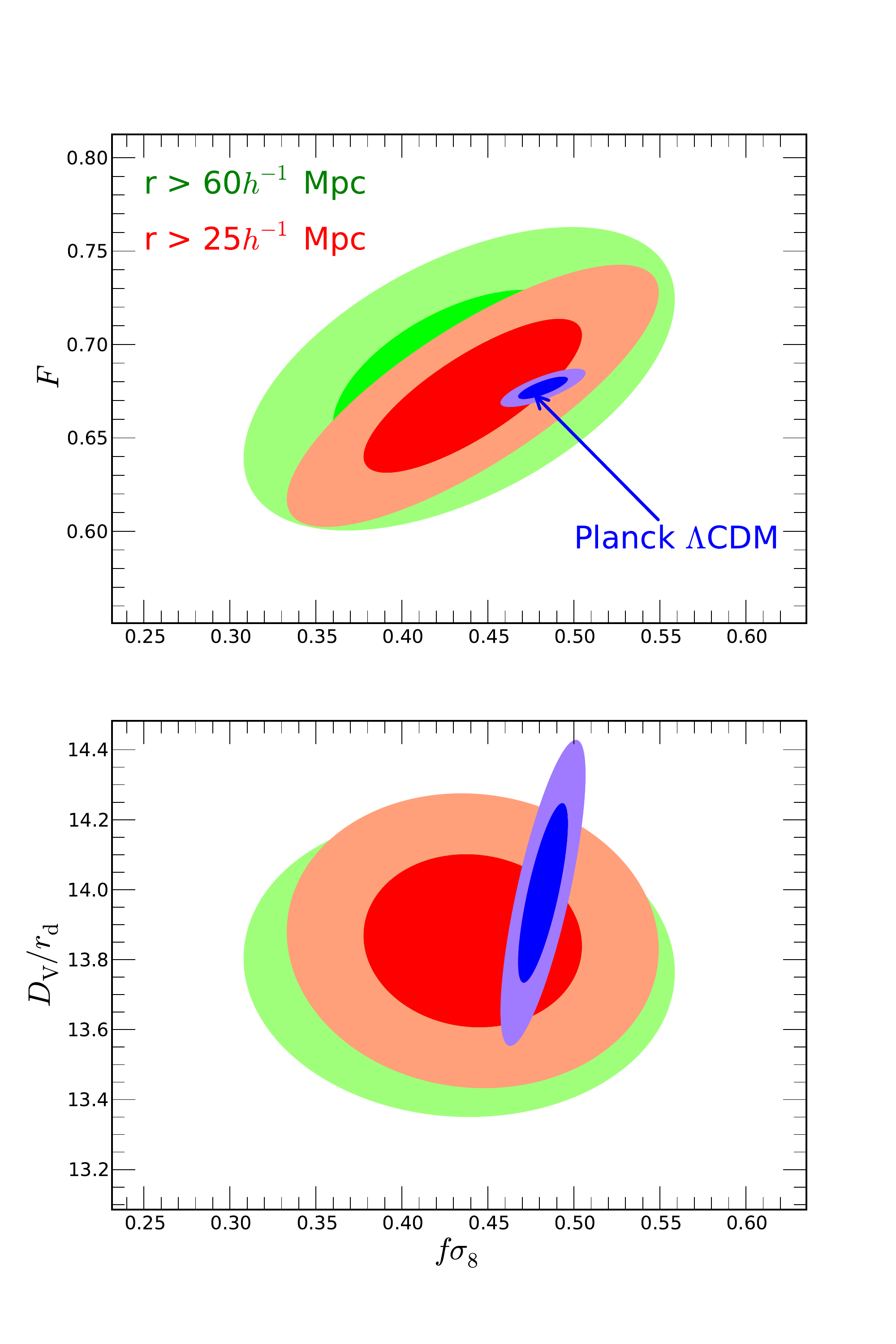}
\caption{Constraints on $f\sigma_8$, $F$ and $D_{\rm V}/r_\mathrm{d}$ from large-scales
only (green contours) and from all scales (red contours), along with predictions
of $\Lambda$CDM-GR normalized by the {\it Planck} results. All estimates are mutually
consistent.}
\label{fig:LS}
\end{figure}

Moments of the correlation function are measured with the best signal-to-noise ratio
at small scales. The model that we use has been tested against numerical
simulations with agreement at the per cent level down to
$r\simeq 25\,h^{-1}$Mpc.  To determine explicitly the contribution of small scales
on our fits, we redo the fit to the monopole and quadrupole keeping only scales
above $r=60\,h^{-1}$Mpc.
The results of this fit and the comparison to the main results are shown in
Fig.~\ref{fig:LS}.

The inclusion of scales between $30\,h^{-1} < r < 60\,h^{-1}$Mpc improves our
constraints by approximately a factor of 2. For the main parameters of
interest we find $b\sigma_8 = 1.28\pm0.07$, $f\sigma_8=0.433\pm0.050$, $D_{\rm
V}/r_\mathrm{d}=13.78\pm0.17$, and $F=0.682\pm0.033$. The biggest improvement is in
the variables that are determined from the amplitudes of the multipoles such as
$b\sigma_8$, $f\sigma_8$ and $F$. Improvement in $D_{\rm V}/r_\mathrm{d}$ is
more modest, because most of the information about this quantity is produced by
the BAO peak in the monopole at $r\simeq 100\,h^{-1}$Mpc.

Small-scale clustering pushes $f\sigma_8$ and $D_{\rm V}/r_\mathrm{d}$ to higher values
and $F$ to lower values. The two estimates, however, are highly consistent, the
$\chi^2$ offset between the two being $\chi^2 = 0.29$ for 3 degrees of freedom.
Both sets of measurements are consistent with {\it Planck} data.  The $\chi^2$
difference between large-scale-only measurements and {\it Planck} inferred
values is 1.8 for 3 degrees of freedom,  while the difference between large-scale-only measurements
and the ones using all scales above 25$h^{-1}$ Mpc is 0.3 for 3 degrees of
freedom.

\section{Cosmological implications}
\label{sec:cosmoconstraints}

In following subsections, we constrain parameters of standard $\Lambda$CDM-GR
model by combining our measurements with the CMB and previous, independent BAO
measurements. We also examine possible deviations from the standard model by
considering phenomenological modifications to both $\Lambda$ and GR parts. 

As a CMB data set we adopt the recent measurements of CMB temperature
fluctuations by the {\it Planck} satellite \citep{Planck} supplemented by
low-$\ell$ measurements of CMB polarization from the {\it WMAP} misssion
\citep{WMAP9} and the high-$\ell$ measurements from the Atacama Cosmology
Telescope \citep[ACT]{ACT} and the South Pole Telescope \citep[SPT]{SPT}. For
the rest of the paper, we will refer to this combination of CMB data as {\it
ePlanck}\footnote{When computing the CMB likelihood we make the same
assumptions as \citet{Planck}.  For example, we assume a minimum neutrino mass
of $\sum m_\nu = 0.06\,{\rm eV}$. This affects the time of matter-radiation
equality and angular-diameter distance to last scattering, as well as early
integrated Sachs--Wolfe effect and the lensing potential.}.  For our BAO data
compilation we use measurements from \citet[$z=0.106$]{6dFBAO},
\citet[$z=0.32$]{DR10BAO}\footnote{We only use a measurement of BAO from the lower redshift 
(LOWZ) sample since the measurement from the CMASS sample is highly correlated
with our own estimate of $D_{\rm V}/r_\mathrm{d}$.} and \citet[$z=0.60$]{WiggleZBAO}. 

To sample cosmological parameter space, we use the Monte Carlo Markov Chain
(MCMC) technique implemented by the \textsc{cosmomc} package \citep{cosmomc}.

\subsection{$\Lambda$CDM-GR}

In a spatially flat $\Lambda$CDM-GR model, the expansion history of the Universe
and the growth of perturbations can be fully described by six parameters. We
choose these to be $\boldsymbol{ p}_{\rm \Lambda CDM}=\left[\Omega_\Lambda, \Omega_{\rm
b}h^2, n_{\rm s}, \sigma_8(0), \tau, H_0\right]$.
The mean values and 1$\sigma$ confidence levels are
listed in Table~\ref{tab:lcdm}.

\begin{table}
\begin{center}
\begin{tabular}{|l|l|l|l|}
\hline
\multicolumn{4}{|c|}{$\Lambda$CDM-GR}\\
\hline
Parameter & \textit{ePlanck} & BOSS + \textit{ePlanck} & BOSS + \textit{ePlanck} + BAO \\
\hline
$100\Omega_{\rm b}h^2$ & $2.21\pm0.03$ & $2.21\pm0.02$& $2.22\pm0.02$\\
$\Omega_\Lambda$ & $0.685\pm0.017$ & $0.692 \pm0.011$& $0.696\pm0.009$\\
$n_{\rm s}$ & $0.960\pm0.007$ & $0.961\pm0.006$& $0.962\pm0.005$\\
$\sigma_8(0)$ & $0.829\pm0.012$ & $0.823\pm0.011$ & $0.821\pm0.011$\\
$100\tau$ & $8.91\pm1.30$ & $8.91\pm1.24$& $9.02\pm1.23$\\
$H_0$ & $67.3\pm1.2$ & $67.8\pm0.84$ & $68.1\pm0.7$\\
\hline
\end{tabular}
\caption{Constraints on basic parameters of $\Lambda$CDM-GR.}
\label{tab:lcdm}
\end{center}
\end{table}

By combining {\it BOSS} DR11 results with {\it Planck} data, we are able
to achieve a 1.6 per cent constraint on $\Omega_\Lambda$, a 1.3 per cent
constraint on $\sigma_8(0)$ and a 1.2 per cent constraint on $H_0$.  After
including BAO data set, the constraints improve to 1.3 per cent on
$\Omega_\Lambda$ and a 1.0 per cent constraint on $H_0$, while relative
constraint on $\sigma_8(z=0)$ does not change. The constraints on $\Omega_\mathrm{b}$
and $n_\mathrm{s}$ are dominated by the information from the \textit{ePlanck} data set.

\subsection{Spatial curvature}

\begin{figure}
\includegraphics[width=\linewidth]{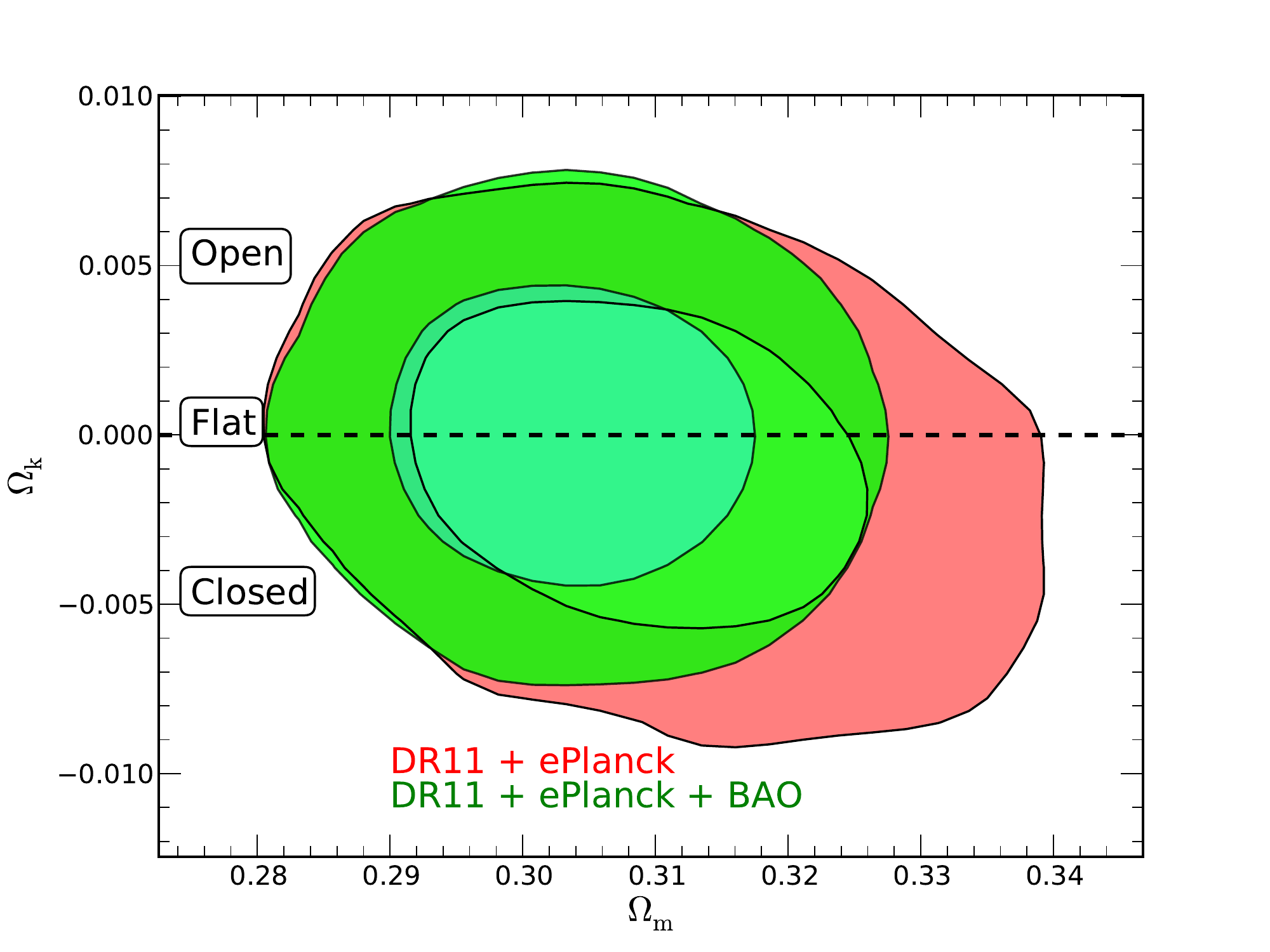}
\caption{Constraints on spatial curvature and nonrelativistic matter
density from the combination of {\it BOSS} DR11 data with CMB and BAO
data sets. The contours correspond to 1$\sigma$ and $2\,\sigma$ confidence levels
in posterior likelihood.}
\label{fig:omokconstraints}
\end{figure}

We now relax the assumption that the spatial curvature is zero and allow the
$\Omega_{\rm k}$ parameter to vary along with $\boldsymbol{ p}_{\rm \Lambda CDM}$.  The
posterior confidence regions on curvature and nonrelativistic matter density
are shown in Fig.~\ref{fig:omokconstraints}.

We find $1 + \Omega_{\rm k}=0.999\pm0.003$ (a 0.3 per cent constraint) when
combining {\it BOSS} DR11 with {\it ePlanck} and $1+\Omega_{\rm
k}=1.000\pm0.003$ (a 0.3 per cent constraint) when adding the BAO compilation.
In both cases the results are perfectly consistent with a spatially flat
Universe.

\subsection{Time dependence of DE}

Alternative models of DE predict a time-dependent EoS $w(z)$. For
a wide range of DE models that do not exhibit sudden transitions or large amount
of DE at early times, for example models based on cosmological scalar fields,
this time-dependence can be adequately parametrized by two parameters 
\begin{equation}
w(z) = w_0 + w_{\rm a}\frac{z}{1+z} 
\end{equation}
\noindent
\citep{wowa, wowaLinder}.
\citep[For DE models that do not belong to this family see e.g.][]{EDE1,
EDE2}. This reduces to our standard model for $w_0 = -1$ and $w_{\rm a} = 0$.

\begin{figure}
\includegraphics[width=\linewidth]{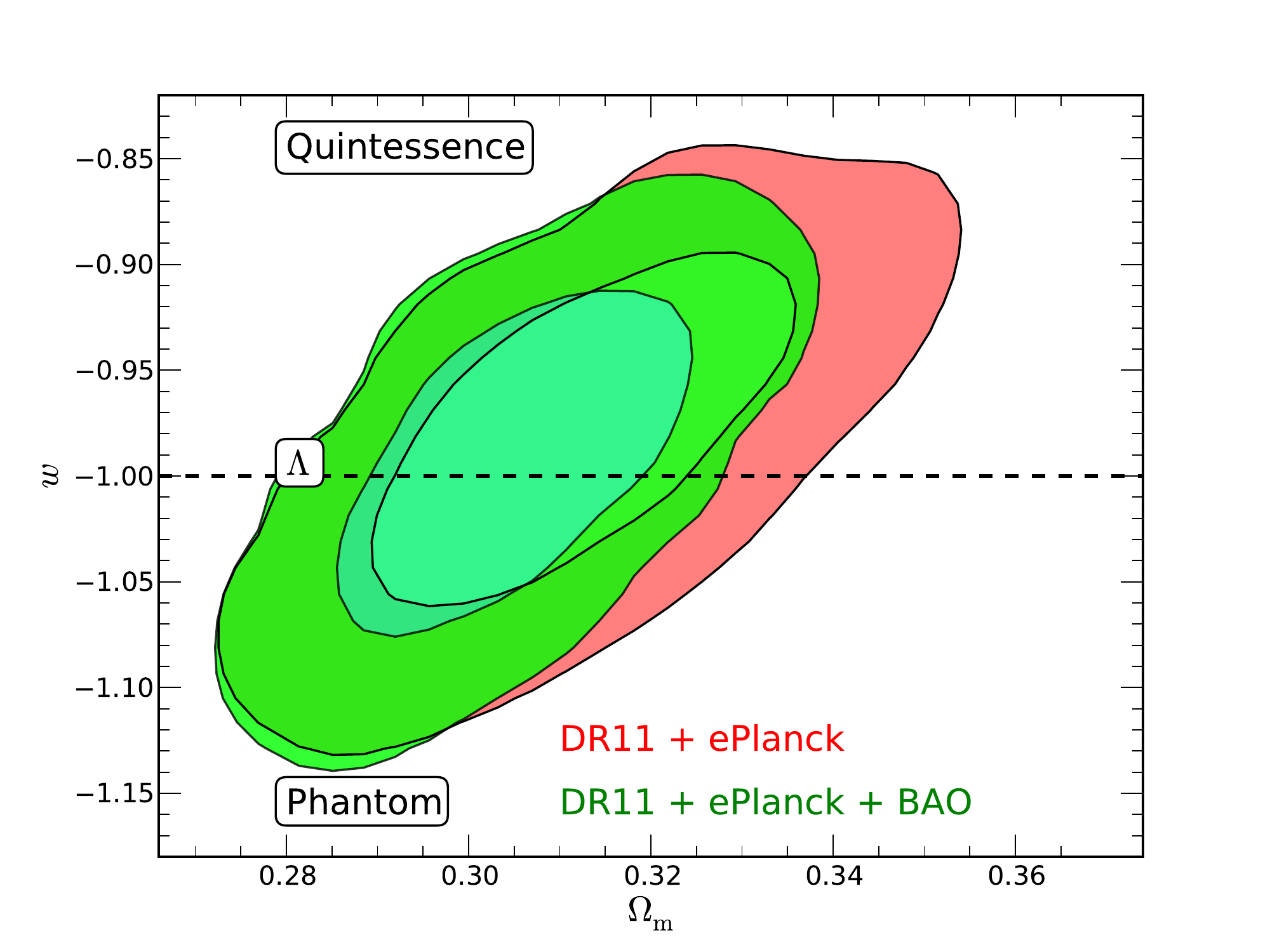}
\caption{Constraints on $w$ and nonrelativistic matter density from the
combination of {\it BOSS} DR11 data with CMB and BAO data sets.
The contours correspond to 1$\sigma$ and $2\,\sigma$ confidence levels in posterior.}
\label{fig:wconstraints}
\end{figure}

We first set $w_{\rm a}$ to zero and check if there is an evidence for 
$w$ to differ from $-1$ on average. The confidence level contours on $w$
and nonrelativistic matter density are shown in Fig.~\ref{fig:wconstraints}.

This analysis yields $w = -0.983 \pm 0.075$ (a 8 per cent constraint) when {\it BOSS} DR11 is
combined with {\it ePlanck} data and $w = -0.993 \pm 0.056$ (a 6 per cent
constraint) when the BAO compilation is added. In both cases, the results are
perfectly consistent with a cosmological constant ($w=-1$). Our constraints on
$w$ differ significantly from the DR9 results presented in
\citet{Samushiaetal2013}, where we detected up to $2\,\sigma$ preference
for $w>-1$. This change is mainly due to two differences.
We now use {\it ePlanck} as our CMB data set, which predicts a higher value
for the non-relativistic matter density.
Also, our new measurements, although consistent with DR9 results, have
moved in the direction that makes them more consistent with the CMB results 
(see Fig.\ref{fig:dvFfs8}).

\begin{figure}
\includegraphics[width=\linewidth]{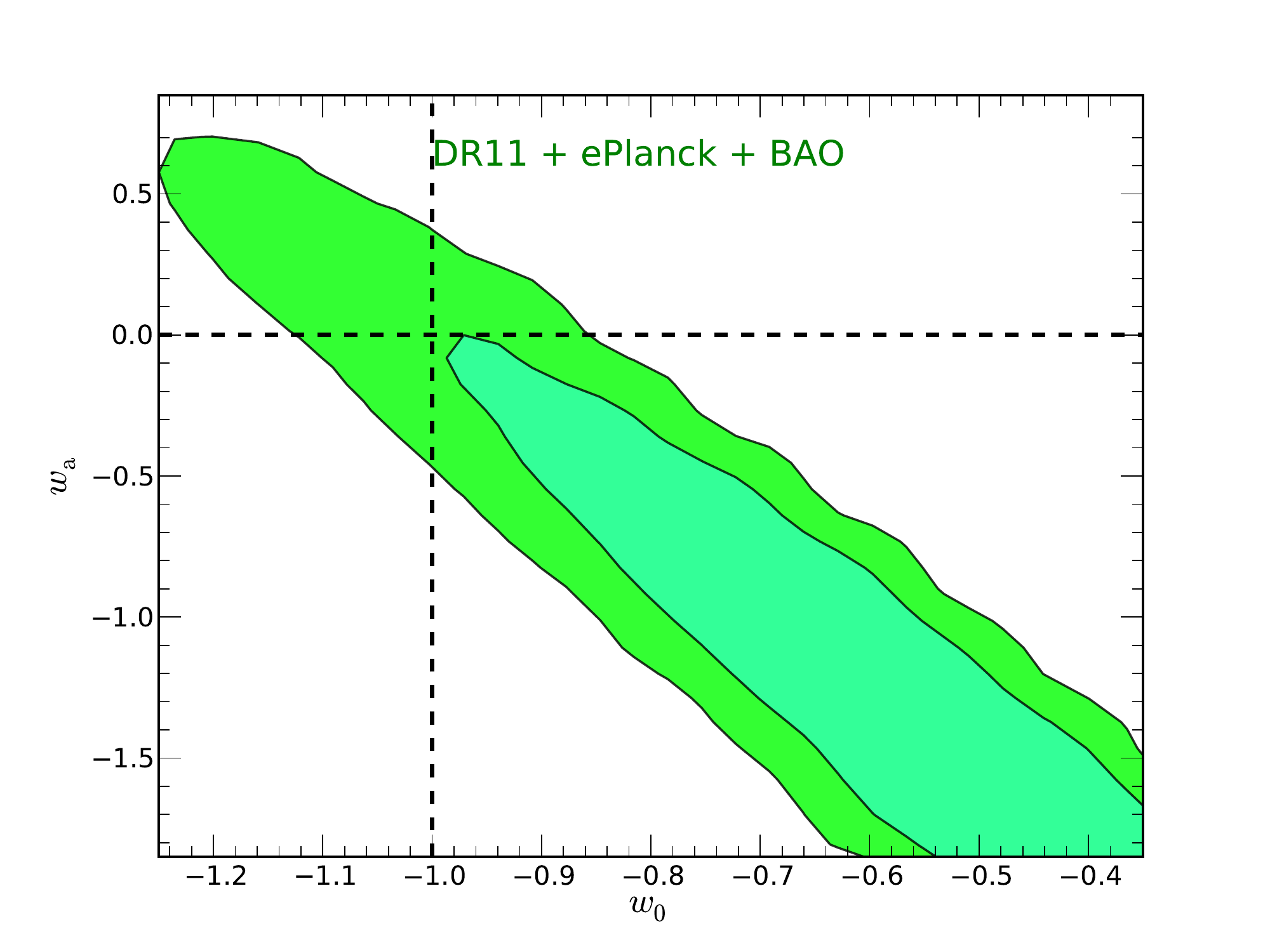}
\caption{Constraints on $w$ and $w_{\rm a}$ from the combination of {\it BOSS}
DR11 data with CMB and BAO data sets. The contours correspond to 1$\sigma$ and
$2\,\sigma$ confidence levels in posterior likelihood. The $\Lambda$CDM prediction is
consistent at the $1.5\,\sigma$ level.}
\label{fig:fom}
\end{figure}

Finally, we consider a model in which the spatial curvature is a free parameter
and both $w_0$ and $w_{\rm a}$ are allowed to vary. The constraints on this
model are shown in Fig.~\ref{fig:fom}.
The DR11 data alone, even after combining with {\it ePlanck}, is not able to break all
the degeneracies of this large parameter space. When DR11 and {\it ePlanck} are
combined with the BAO, we see a preference for larger values of $w_0$ and smaller
values of $w_{\rm a}$.
The $\Lambda$CDM value of $w_0 = -1$ and $w_{\rm a}=0$, however, is still within the $2\,\sigma$
confidence level.

\subsection{Deviations from GR}

MG predict scale dependence of bias and growth
rate even in the linear regime and the effect of small-scale screening
mechanisms is difficult to model. This makes devising a completely
self-consistent test of MG models a non-trivial task. A number of proposals for
parametrizing families of MG models have been discussed recently \citep[see
e.g.][]{MG1, MG2, MG4, MG3}. These parametrizations, however, are difficult to
correctly implement in practice for a few reasons. First, they rely on the linear
theory and are not expected to work below scales of $\sim 25\,h^{-1}$Mpc.
Secondly, they require a large number of free parameters and such a large
parameter space cannot be effectively constrained by current data.  For these
reasons, we follow the approach of \citet{Samushiaetal2013} and apply several
few-parameter consistency tests to our measurements. 

We parametrize the growth rate as a function of $\Omega_{\rm m}$ using
\begin{equation}
f = \left[\frac{\Omega_{\rm m}(z)}{E(z)}\right]^\gamma
\end{equation}
\noindent
\citep{gamma}. This approach does not provide a fully self-consistent test of
MG models, as MG models predict a more complex change in observables
compared to GR. This parametrization is, however, easy to implement and provides a simple
consistency test. In GR, we expect the $\gamma$-index to be equal to 0.554. Measuring
a significantly higher value would indicate a preference for a force weaker than GR
gravity and vice versa. In our fits, we apply a hard prior of $\gamma < 1.0$.

When constraining deviations from GR, we fix DE to be a cosmological constant. We
also ignore the CMB power spectrum on large scales ($\ell < 50$) to ensure that
CMB data are used only to constrain the background evolution. The
parametrizations that we use are not physically motivated and are simply meant
to describe effective gravity at low redshifts rather than provide a full model
that works accurately at all redshifts up to last-scattering surface.

\begin{figure}
\includegraphics[width=\linewidth]{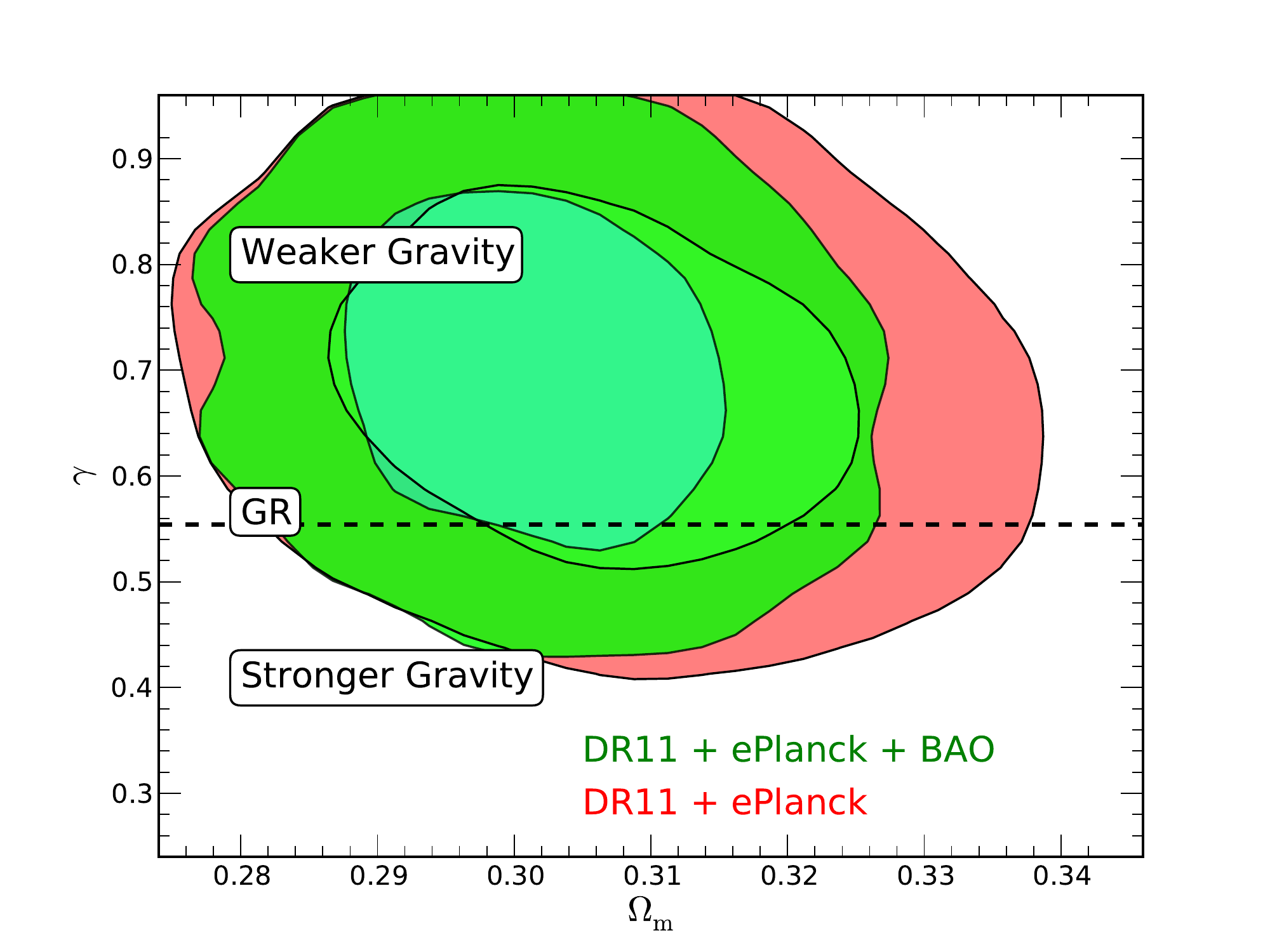}
\caption{Constraints on $\gamma$ index and nonrelativistic matter density from
the combination of {\it BOSS} DR11 data with CMB, SNIa, BAO and $H_0$ data sets.
The contours correspond to 1$\sigma$ and $2\,\sigma$ confidence levels in posterior
likelihood. The best fit is consistent with GR at $1\,\sigma$.}
\label{fig:gammaconstraints}
\end{figure}

Constraints on $\gamma$ and $\Omega_{\rm m}$ are shown in
Fig.~\ref{fig:gammaconstraints}.  When combining {\it BOSS} DR11 with {\it
ePlanck} data, we recover $\gamma = 0.691 \pm 0.111$ (a 16 per cent
measurement). With the BAO data set, we recover $\gamma = 0.699 \pm 0.110$ (a 16
per cent measurement).  The values are within $1.2\,\sigma$ confidence of GR
values but favour a weaker gravity.

\begin{figure}
\includegraphics[width=\linewidth]{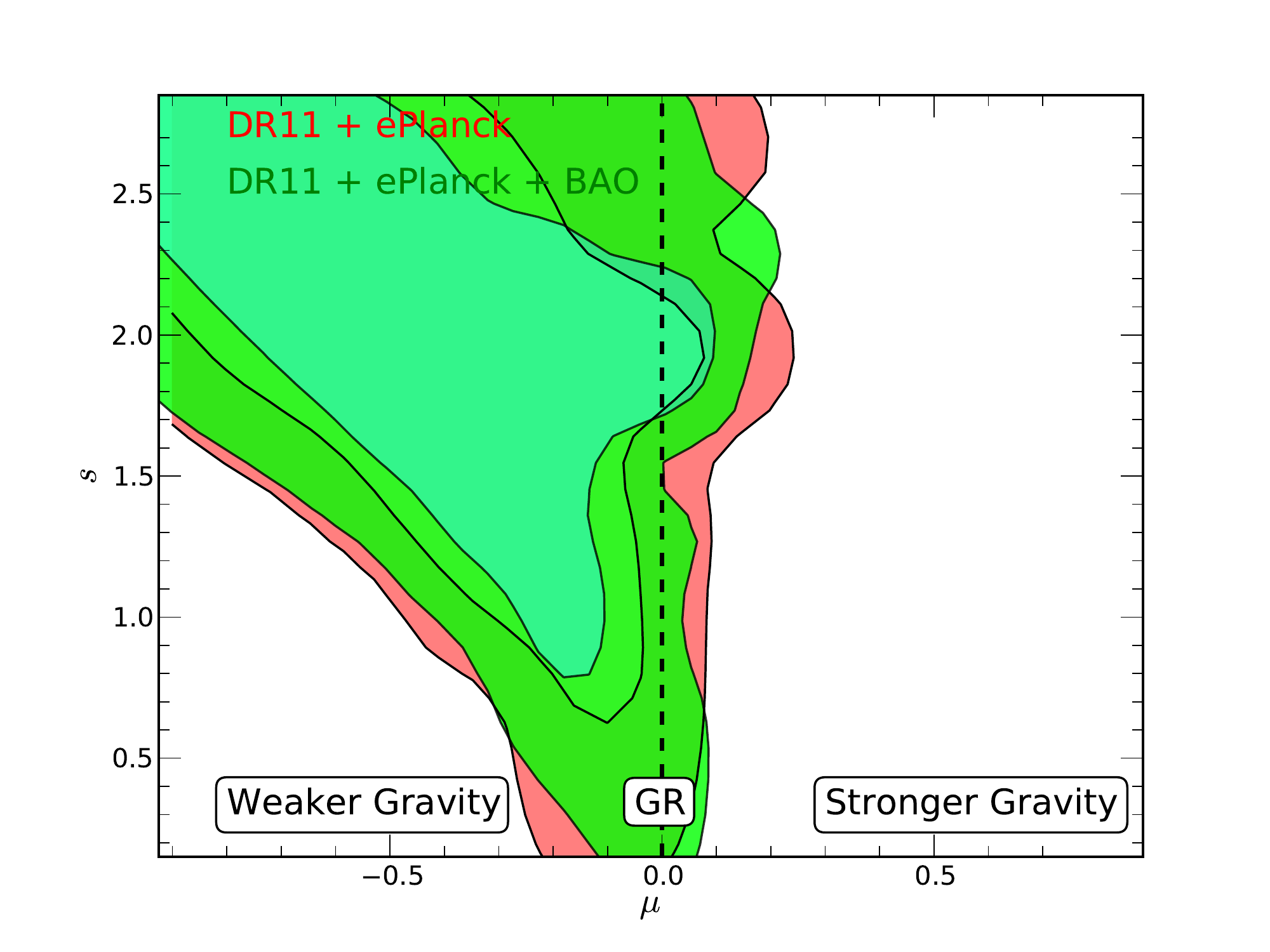}
\caption{Constraints on $\mu$ and $s$ from the combination of {\it BOSS} DR11
data with CMB and BAO data sets. The contours correspond to 1$\sigma$ and
$2\,\sigma$ confidence levels in posterior likelihood.}
\label{fig:musconstraints}
\end{figure}

Next, we parametrize the linear equation of growth following the approach of 
\citet{MG5} as
\begin{equation}
\ddot{\delta} + \left(2 + \dot{H}\right)\dot{\delta} = \frac{3}{2}\Omega_{\rm m}(z)G\delta\left(1 + \mu a^s\right),
\end{equation}
\noindent
where $\delta$ is a matter overdensity, the overdot denotes a derivative with
respect to $\ln a$, $G$ is the gravitational constant, and $\mu$ and $s$ are
parameters describing deviations from GR. The GR limit is recovered when $\mu = 0$,
where negative values of $\mu$ correspond to weaker than GR gravity and vice versa.
The $s$ parameter dictates how rapidly the modifications are set larger values of $s$ corresponding
to the modifications that appear at later times. Since large values of $s$ correspond
to models in which gravity is indistinguishable from GR until some low redshift when
the modification suddenly becomes significant, they are basically unconstrained. We place a flat prior of
$0 < s < 3$ to avoid this problem.
The confidence level contours of $\mu$ and $s$ are shown in Fig.~\ref{fig:musconstraints}.

The GR predictions are within $2\,\sigma$ in posterior likelihood. Similar to
$\gamma$-parametrization, the data again provide a mild preference for a weaker
than GR gravity. This result is consistent with the DR9-based results
reported in \citet{Samushiaetal2013}.

\section{Conclusions}
\label{sec:conclusions}

We have used the anisotropic clustering of galaxies in the {\it BOSS} DR11 data
set to simultaneously constrain the growth rate, the redshift--distance
relationship and the expansion rate at the redshift of $z=0.57$.  Overall, our
measurements are in good agreement with the results of the {\it Planck}
satellite propagated to low redshifts assuming $\Lambda$CDM-GR.

By combining our measurements of $f$, $D_{\rm V}$ and $F$ with the CMB data we
were able to derive tight constraints on basic cosmological parameters and
parameters describing deviations from the $\Lambda$CDM-GR model. We were able to
constrain the curvature of Universe with 0.3 per cent precision, the DE EoS
parameter $w$ with 8 per cent precision and the $\gamma$-index for growth
with 16 per cent precision.

When we vary the background expansion within $\Lambda$CDM predictions of the
{\it Planck} data we measure the growth rate (parametrized by $\gamma$)  to be
weaker but consistent within $1.2\,\sigma$ of GR predictions. This preference
for lower values of growth rate has also been observed in other similar
low-redshift measurements \citep[see e.g.][for discussion]{macaulay}. Our
measurement of $f\sigma_8$ follows this trend but is closer to the GR
predictions compared to the DR9 results of \citet{Reidetal2012} and the DR11
measurement of \citet{Beutler2013}.

Similar measurements from a lower redshift (LOWZ) sample of {\it BOSS} galaxies
will provide a complementary measurement of the growth rate in the DE-dominated
redshift range of $0.2 < z < 0.43$, which will significantly strengthen the
constraining power over possible GR modifications and can potentially increase
the significance of the `low growth rate' signal.

\section*{Acknowledgements}

LS gratefully acknowledges support by the European Research Council. BAR
gratefully acknowledges support provided by NASA through Hubble Fellowship
grant 51280 awarded by the Space Telescope Science Institute, which is operated
by the Association of Universities for Research in Astronomy, Inc., for NASA,
under contract NAS 5-26555.  Funding for SDSS-III has been provided by the
Alfred P.  Sloan Foundation, the Participating Institutions, the National Sci-
ence Foundation, and the U.S.  Department of Energy Office of Sci- ence. The
SDSS-III web site is http://www.sdss3.org/. 

SDSS-III is managed by the Astrophysical Research Consortium for the
Participating Institutions of the SDSS-III Collaboration including the
University of Arizona, the Brazilian Participation Group, Brookhaven National
Laboratory, University of Cambridge, CarnegieMellon University, University of
Florida, the French Participation Group, the German Participation Group,
Harvard University, the Instituto de Astrofisica de Canarias, theMichigan
State/Notre Dame/JINA Participation Group, Johns Hopkins University, Lawrence
Berkeley National Laboratory, Max Planck Institute for Astrophysics, Max Planck
Institute for Extraterrestrial Physics, New Mexico State University, New York
University, Ohio State University, Pennsylvania State University, University of
Portsmouth, Princeton University, the Spanish ParticipationGroup, University of
Tokyo, University of Utah, Vanderbilt University, University of Virginia,
University of Washington, and Yale University. 

We acknowledge the use of MCMC data from Planck Legacy Archive \\(\url{http://www.sciops.esa.int/index.php?project=planck&page=Planck_Legacy_Archive}).

Numerical computations were done on the Sciama High Performance Compute 
cluster which is supported by the ICG, SEPNet and the University of Portsmouth.

\appendix

\onecolumn

\section{CMB shape prior}

\begin{figure}
\includegraphics[width=\linewidth]{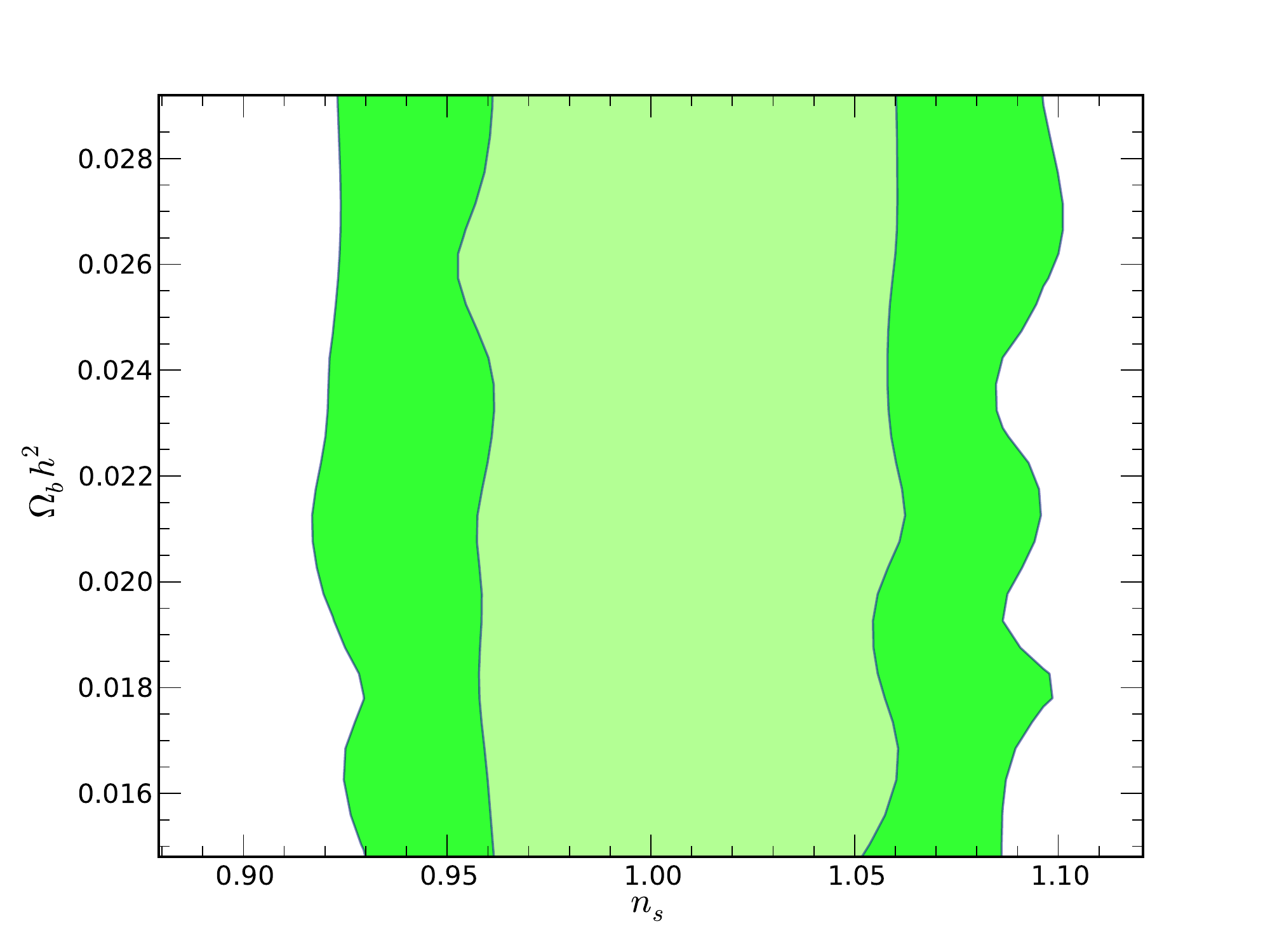}
\caption{Posterior likelihood in $\textbf{p}_{\rm sh}$ from {\it BOSS} {\it DR11} data
only. The $\Omega_{\rm b}$ remains unconstrained in a 10$\sigma$ range around
the CMB measurement.}
\label{fig:gcshape}
\end{figure}

Figure \ref{fig:gcshape} displays the posterior likelihood of $p_{\rm sh}$
obtainable from {\it DR11} data alone.

The likelihood surface does not close even within $\pm 10 \sigma$ of the CMB
constraints. Previous studies either fix the shape parameters to their CMB
best-fit values \cite[e.g.][]{DR7RSD}, let $\Omega_{\rm m}h^2$ vary and fix
the rest \citep[e.g.][]{WiggleZRSD} , or marginalise over them by taking a
prior centered around CMB best-fit values \citep[e.g.][]{Chuang2012}. 

We adopt a different approach and apply the CMB shape prior to our galaxy
clustering likelihood. Since later we will combine our results with Planck
data to obtain constraints on DE and MG parameters one may be led to an erroneous
impression that the CMB data is being double-counted. We demonstrate below that
this is not the case.

Let $\mathcal{L}^C(a, b)$ be a CMB likelihood, where $a$ are shape parameters
and $b$ are other parameters that may be related to DE and gravity parameters
of interest [$b(w, \gamma, \ldots)$].  Let $\mathcal{L}^G(a, c)$ be galaxy
likelihood, where $c$ are DE and gravity dependent [$c(w, \gamma, \ldots)$].  For
simplicity, assume $a$, $b$ and $c$ to be scalars and that all
likelihoods are multivariate Gaussian.

In our approach we take a CMB shape prior 
\begin{equation}
\displaystyle\int \mathcal{L}^C(a,b)db
\end{equation}
apply it to galaxy data
\begin{equation}
\displaystyle\int \mathcal{L}^C(a,b)L^G(a,c)dbda
\end{equation}
and then combine it with full CMB likelihood
\begin{equation}
\mathcal{L}_1(b'(x)c(x))=\displaystyle\int \mathcal{L}^C(a,b)\mathcal{L}^G(a,c)\mathcal{L}^C(a',b')dbdada'
\end{equation}
where $x = (w, \gamma, \ldots)$.

Let's compare this expression to that produced by directly combining the two
likelihoods
\begin{equation}
\mathcal{L}_2(b(x)c(x))=\displaystyle\int \mathcal{L}^C(a,b)\mathcal{L}^G(a,c)da
\end{equation}

$\mathcal{L}_1(b, c)$ and $\mathcal{L}_2(b, c)$ are also Gaussian with
\begin{eqnarray}
\sigma_{1b}^2 &=& \sigma_{Cb}^2\\
\sigma_{1c}^2  &=& \frac{\sigma_{Ca}^2+\sigma_{Ga}^2(1-r_{Gac})}{\sigma_{Ca}^2+\sigma_{Ga}^2}\sigma_{Gc}^2\\
\rho_{1bc}        &=& 0
\end{eqnarray}
\begin{eqnarray}
\nonumber
\sigma_{2b}^2 &=& \frac{(\sigma_{Ca}^2(1-r_{Cab})+\sigma_{Ga}^2)(\sigma_{Ca}^2(1-r_{Cab})+\sigma_{Ga}^2(1-r_{Gac}))}{\sigma_{Ca}^4(1-r_{Cab})+\sigma_{Ca}^2\sigma_{Ga}^2(2-r_{Gac}^2-r_{Cab}^2-3r_{Cab}^2r_{Gac}^2)+\sigma_{Ga}^4(1-r_{Gab})}\sigma_{Cb}^2\\
\nonumber
\sigma_{2c}^2 &=& \frac{(\sigma_{Ca}^2+\sigma_{Ga}^2(1-r_{Gac}))(\sigma_{Ca}^2(1-r_{Cab})+\sigma_{Ga}^2(1-r_{Gac}))}{\sigma_{Ca}^4(1-r_{Cab})+\sigma_{Ca}^2\sigma_{Ga}^2(2-r_{Gac}^2-r_{Cab}^2-3r_{Cab}^2r_{Gac}^2)+\sigma_{Ga}^4(1-r_{Gab})}\sigma_{Gc}^2\\
\nonumber
\rho_{2bc}        &=&  \frac{r_{Gac}r_{Cab}\sigma_{Ca}\sigma_{Ga}\sigma_{Cb}\sigma_{Gc}(\sigma_{Ca}^2(1-r_{Cab})+\sigma_{Ga}^2(1-r_{Gac}))}{\sigma_{Ca}^4(1-r_{Cab})+\sigma_{Ca}^2\sigma_{Ga}^2(2-r_{Gac}^2-r_{Cab}^2-3r_{Cab}^2r_{Gac}^2)+\sigma_{Ga}^4(1-r_{Gab})}
\end{eqnarray}

$\mathcal{L}_2$ always encloses $\mathcal{L}_1$. To first order in $\sigma_{Ca}/\sigma_{Ga}$
\begin{eqnarray}
\sigma_{1b}^2 &=& \sigma_{Cb}^2\\
\sigma_{1c}^2  &=& (1-r_{Gac})\sigma_{Gc}^2\\
\rho_{1bc}        &=& 0
\end{eqnarray}

\begin{eqnarray}
\sigma_{1b}^2 &=& \sigma_{Cb}^2\\
\sigma_{1c}^2  &=& (1-r_{Gac})\sigma_{Gc}^2\\
\rho_{1bc}        &=& r_{Gac}r_{Cab}\sigma_{Cb}\sigma_{Gc}\frac{\sigma_{Ca}}{\sigma_{Cb}}
\end{eqnarray}

This demonstrates that direct combination of galaxy clustering and CMB data always
produce stronger constraints on derived parameters and therefore the galaxy clustering measurements
obtained by assuming a CMB prior on the shape can be combined with the CMB data without double counting
the information.

\label{lastpage}

\end{document}